\documentclass[11pt,a4paper]{article}
\usepackage{jheppub}

\newcommand{\mt}[1]{\textrm{\tiny #1}}

\newcommand{\sac}{\, , \qquad}
\newcommand{\gym}{g_\mt{YM}}
\newcommand{\eqn}[1]{(\ref{#1})}

\usepackage{epsfig}

\def\N{{\cal N}}

\def\Tr{{\rm Tr}}

\def\det{{\rm det}}

\newcommand{\Dslash}{D\mkern-11.5mu/\,} %% generalized Dirac operator
\newcommand{\delslash}{{\partial\mkern-9mu/}}

\def\Dbarslash{\,\,{\raise.15ex\hbox{/}\mkern-12mu {\bar D}}}
\def\Dslash{\,\,{\raise.15ex\hbox{/}\mkern-12mu D}}
\def\delslash{\,\,{\raise.15ex\hbox{/}\mkern-9mu \partial}}
\def\delbarslash{\,\,{\raise.15ex\hbox{/}\mkern-9mu {\bar\partial}}}

\newcommand{\EQ}[1]{\begin{equation} #1 \end{equation}}

\newcommand{\SP}[1]{\begin{equation}\begin{split} #1
\end{split}\end{equation}}

\newcommand{\beq}{\begin{equation}}
\newcommand{\eeq}{\end{equation}}
\newcommand{\beqs}{\begin{eqnarray}}
\newcommand{\eeqs}{\end{eqnarray}}
\newcommand{\lsim}{\mathrel{\raisebox{-
.6ex}{$\stackrel{\textstyle<}{\sim}$}}}
\newcommand{\gsim}{\mathrel{\raisebox{-
.6ex}{$\stackrel{\textstyle>}{\sim}$}}}

\def\Dslash{\raisebox{1pt}{$\slash$} \hspace{-9pt} D}
\def\hbar{\hspace{0pt}\raisebox{1pt}{$-$} \hspace{-7pt} h}

\def\di{\mbox{d}}

\newcommand{\be}{\begin{equation}}
\newcommand{\ee}{\end{equation}}
\newcommand{\bea}{\begin{eqnarray}}
\newcommand{\eea}{\end{eqnarray}}

\newcommand{\rc}{\nonumber\\}
\newcommand{\bear}{\begin{eqnarray}}
\newcommand{\eear}{\end{eqnarray}}

\title{Towards holographic walking from ${\cal N}=4$ super Yang-Mills}
\author[a]{S.~Prem Kumar,}
\author[b,c]{David Mateos,}
\author[c]{Angel Paredes}
\author[a]{and Maurizio Piai}
\affiliation[a]{Department of Physics, 
Swansea University, 
Singleton Park, Swansea, SA2 8PP, UK.}
\affiliation[b]{Instituci\'o Catalana de Recerca i Estudis Avan\c cats (ICREA), Passeig Llu\'\i s Companys 23, E-08010, Barcelona, Spain} 
\affiliation[c]{Departament de F\'\i sica Fonamental \&  Institut de 
Ci\`encies del Cosmos (ICC), Universitat de Barcelona (UB), Mart\'{\i}  i Franqu\`es 1, E-08028 Barcelona, Spain.}  

\date{\today}

\abstract{We propose that a holographic description of `walking'
 behaviour, namely quasi-conformal dynamics relevant for technicolor
 models, can be obtained from relevant deformations of ${\cal N}=4$
 super Yang-Mills. We consider deformations which drive the theory
 close to the ${\cal N}=1$ Leigh-Strassler fixed point, eventually
 deviating from it in the deep IR. We use the Pilch-Warner dual
 supergravity description of the flow between the ${\cal N}=4$ and the
 ${\cal N}=1$ fixed points to focus on observables that  
 only require knowledge of the walking
  region. These include large anomalous dimensions of quark bilinear
  operators, which we study via probe D7-branes. We also make a first
  attempt at describing the theory beyond the walking region by
  introducing an infrared cut-off, in the spirit of hard-wall
  models. In this case we find a light, dilaton-like scalar state, but
  whether this mode persists in the exact theory remains an open
  question.} 
  
\keywords{D-branes, Supersymmetry and Duality, 1/N Expansion,
  Gauge-gravity correspondence, QCD Phenomenology} 

\emailAdd{s.p.kumar@swansea.ac.uk} 
\emailAdd{dmateos@icrea.cat} 
\emailAdd{aparedes@ffn.ub.es} 
\emailAdd{m.piai@swansea.ac.uk} 

\begin{document}

\begin{flushright}
ICCUB-10-203
\end{flushright}

\maketitle
\setlength{\parskip}{8pt}
%\newpage
\section{Introduction}
\label{intro}

Walking technicolor (WTC)~\cite{WTC} is an appealing candidate for dynamical electro-weak symmetry breaking (EWSB) (see~\cite{reviews,P} for reviews on the topic). The basic idea is that EWSB may be triggered by the formation of a non-trivial condensate due to a new strongly-coupled interaction, along the  lines of technicolor (TC)~\cite{TC}. What makes WTC special is that, in addition to being strongly coupled, the underlying dynamics is assumed to be approximately conformal over a range of energies above the electro-weak scale, and in this range large anomalous dimensions are present. This improves on some of the difficulties of traditional TC models, as we review in Section \ref{WTC}. 
In view of the recently-started LHC program, it
is important to identify clear experimental signatures distinguishing
this type of proposal from weakly-coupled theories, such as the
minimal version of the Standard Model or its supersymmetric
extensions.  

Unfortunately, the strongly-coupled nature of WTC makes its study by
conventional field theory methods (based on perturbation theory)
difficult, and many non-trivial field theory
questions remain open. One important such question is whether in a WTC
model, due to the spontaneous breaking of dilatation symmetry induced
by the EWSB condensate, there exists a light scalar composite state
with the properties of a light dilaton. If so, this field would have a
phenomenology very similar to that of the light Higgs in the minimal
version of the standard
model~\cite{dilaton,dilatonpheno,dilaton2,dilaton5D}. 

\begin{figure}
\begin{tabular}{cc}
%\,\,\,\,\,\,\,\,\,\,\,\,
\includegraphics[width=0.5 \textwidth]{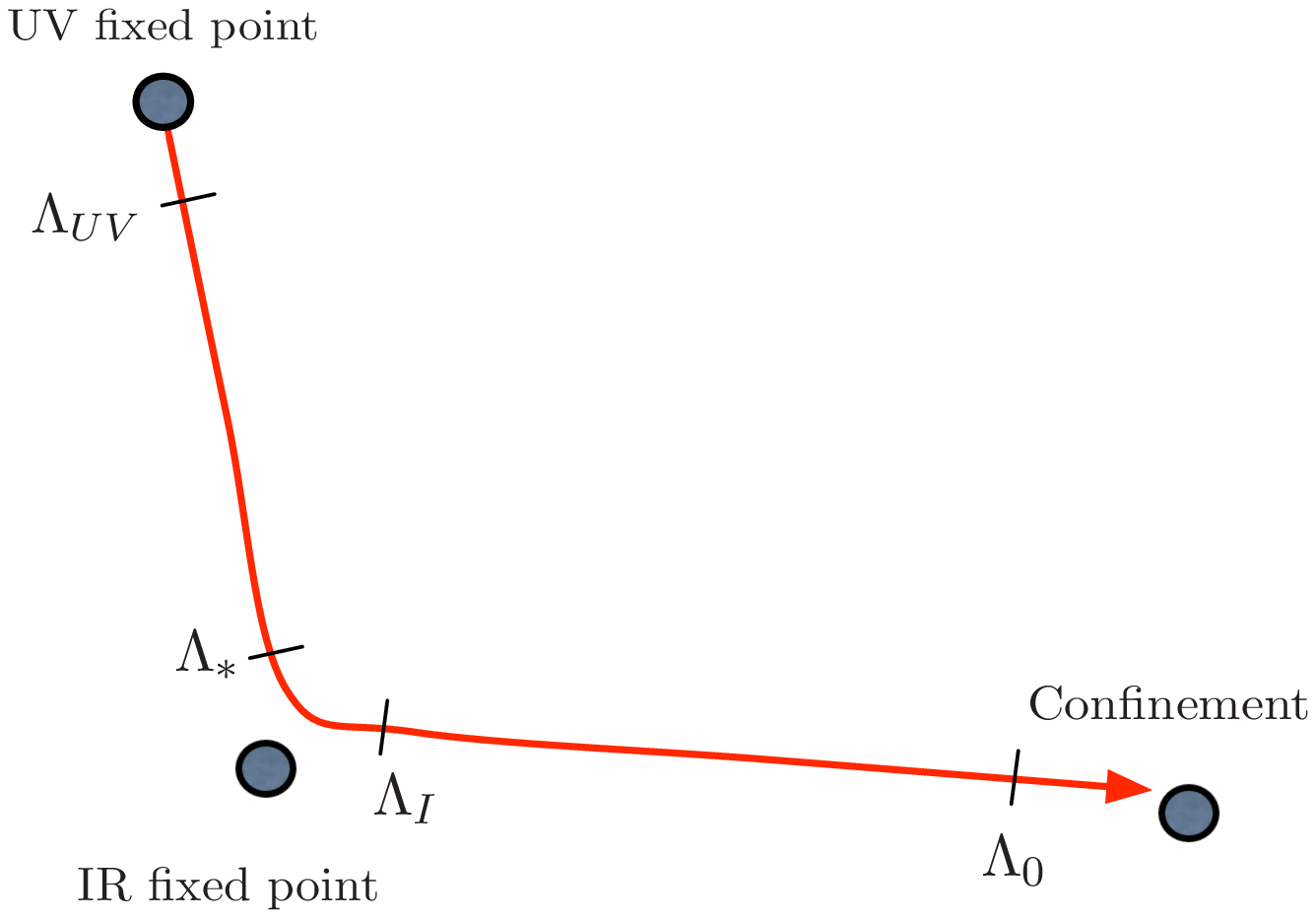} 
%\put(-40,33){$M$}
%\put(-40,5){\small Confinement}
%\put(-125,40){$\Lambda_I$}
%\put(-155,45){$\Lambda_{\ast}$}
%\put(312,20){$m_3$}
&
%\,\,\,\,\,\,\,\,\,\,\,\,
\includegraphics[width=0.5 \textwidth]{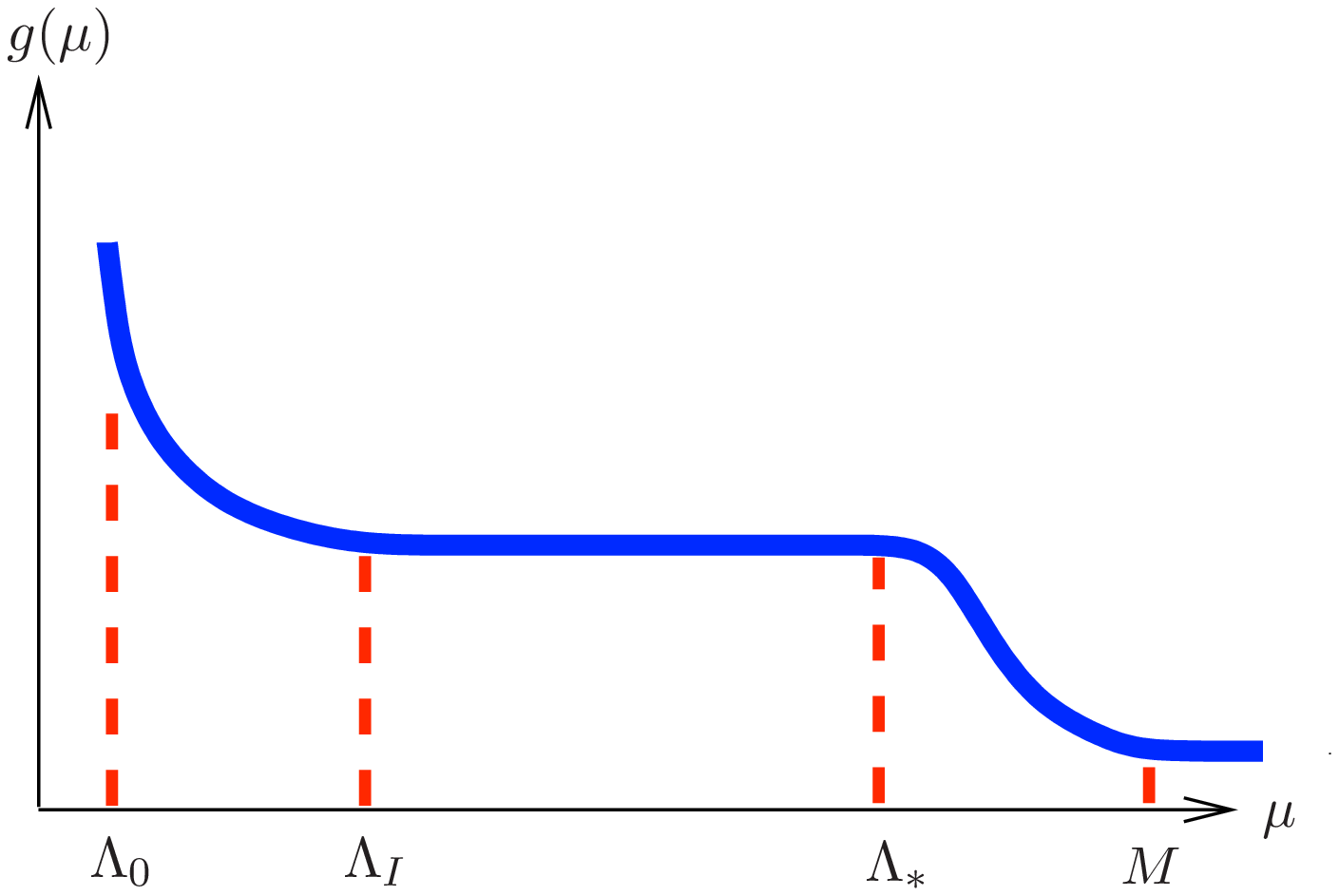} 
%\put(-200,50){$g(\mu)$}
\end{tabular}
\caption{Left: RG flow of a walking  theory.
Right: Cartoon of the effective gauge coupling $g$ as a function of
the renormalization scale $\mu$.}  
\label{Fig:coupling}
\end{figure}

Aside from its potential phenomenological implications, the question
about the existence of a light dilaton is of intrinsic interest from a
purely field-theoretical viewpoint. Indeed, consider a
four-dimensional quantum field theory whose renormalization group (RG)
flow is characterized by three dynamically generated scales
$\Lambda_0<\Lambda_{I}<\Lambda_{\ast}$, as sketched in figure
\ref{Fig:coupling}, with the the following properties. 
The theory is UV-complete, in the sense that at energies above some
high scale $\Lambda_{UV}$ the dynamics is governed by a (possibly
interacting) UV 
fixed point, from which the flow is driven away by some relevant
operator. For example, in  pure Yang-Mills theory the gauge coupling
itself represents a (marginally) relevant deformation driving the
theory away from the UV fixed point, which is free in this
case. Following the RG flow towards lower energies, there exists
another fixed point, which we will refer to as an IR fixed point. The
flow is attracted towards this fixed point at energies below
$\Lambda_{\ast}$, but does not reach the fixed point and is driven
away at energies below  
$\Lambda_I$. In the energy region $\Lambda_I < E < \Lambda_{\ast}$ the
theory is almost scale-invariant and is approximately described by the
conformal field theory (CFT) living at the IR fixed point; in
particular, coupling constants stop running almost
completely. Generically, the operators in this CFT have dimensions
completely different from those at the UV fixed point, and hence in
matching the descriptions along the flow one will find that large,
non-perturbative anomalous dimensions have emerged. Note that, despite
the appearance of figure \ref{Fig:coupling}(left), if the flow comes
very close to the IR fixed point then a large hierarchy $\Lambda_{I}
\ll \Lambda_{\ast}$ develops, as shown in figure
\ref{Fig:coupling}(right). There may or may not be a {\it tunable}
parameter that controls the size of this hierarchy. Below  
$\Lambda_I$ the flow just drifts away from the IR fixed point  and the
theory ultimately confines at some lower scale $\Lambda_0$. With some
abuse of terminology, in the following we will call a theory with
these properties a `walking theory' --- walking technicolor is an
example of such a theory, in which at some scale $\Lambda_W\in
[\Lambda_0,\Lambda_I]$ a condensate forms, spontaneously breaking the
$SU(2)_L\times U(1)_Y$ symmetry of the Standard Model. 

In a scenario like the above, one question we would eventually like to address
is: Under what conditions, for example on the anomalous dimensions of
operators at the IR fixed point, does the spectrum contain a scalar
composite state with the quantum numbers and the couplings of a
dilaton and a mass parametrically lighter than the mass gap
$\Lambda_0$. In order to pose this question, one first needs to
construct a model with the properties of a walking theory, and in
order to address the question one then needs to find an appropriate
tool to study the strong coupling dynamics. In this paper we propose
that certain relevant deformations of  ${\cal N}=4$ super Yang-Mills
(SYM) theory can provide such models, and that their dual string
descriptions provide such tools, at
least in principle. 

Previous steps towards the construction of a walking theory in the
context of the gauge/string duality~\cite{AdSCFT} (see
\cite{reviewAdSCFT} for a review) were given in \cite{NPP,NPR,ENP}. By
wrapping D5-branes on a two-sphere as in \cite{MN}, these references
constructed supergravity solutions that exhibit some of the features
expected of a walking theory. In particular, a suitably-defined gauge
coupling exhibits slow running in a strongly-coupled IR region, and
even more deeply in the IR the theory confines~\cite{NPR}. Moreover,
ref.~\cite{ENP} found evidence of the presence in the spectrum of an
anomalously light scalar with a mass suppressed by the length of the
walking region. The calculation of the spectrum was greatly
facilitated by the use of a gauge-invariant formalism \cite{BHM}.

Despite the appealing features of these models, two obstacles are
encountered when attempting to establish the properties of the light
scalar that are necessary for its identification as a light dilaton. 
First, although in the walking region the gauge coupling
and many other dynamical quantities become effectively constant, the
background metric is not approximately AdS, so a direct connection
with some approximate conformal symmetry is not apparent. Second, a
direct calculation of the couplings of the light scalar is difficult
because in the far-UV the geometry is not asymptotically AdS 
either, and hence the rigorous procedure of holographic
renormalization cannot be straightforwardly applied.

We propose that these two difficulties can be overcome in principle by
constructing a walking theory as a relevant deformation of ${\cal N}=4$
SYM. From the above discussion it is evident that the main necessary
ingredient is the existence of a flow between two conformal field
theories, where the IR fixed point theory is strongly
interacting. In addition, we require that the flow between the two
CFT's have a dual supergravity description which provides the
requisite tool for extracting quantitative results at strong
coupling. The simplest such flow is the one from ${\cal N}=4$ SYM to
the ${\cal N}=1$ supersymmetric Leigh-Strassler fixed point (LS FP). 
In ${\cal N}=1$ language, the ${\cal N}=4$ theory contains a vector
superfield and three chiral superfields $\Phi_i$. When 
deformed by a relevant operator corresponding to a mass $m_3$ for one of
these chiral superfields, say $\Phi_3$, it is known that the ${\cal N}=4$ theory
flows to a strongly interacting ${\cal N}=1$ SCFT, first identified by
Leigh and Strassler \cite{LS}. 

The five-dimensional supergravity solution dual to the LS FP was found
in \cite{KPW} 
(see also \cite{5d}). The ten-dimensional supergravity solution describing the
entire LS flow between the UV and the IR fixed points was found by
Pilch and Warner \cite{PW} by uplifting the five-dimensional flow of
\cite{FGPW}. 
As we will
review, the PW flow is described by a completely smooth supergravity
solution that is known in closed form for all practical purposes. 

\begin{figure}
\begin{center}
\includegraphics[width=0.75 \textwidth]{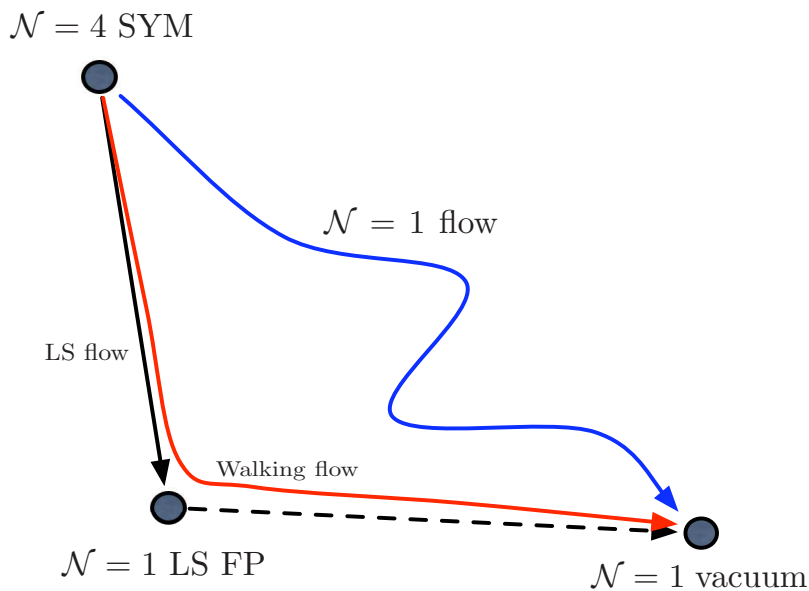} 
\caption{RG flows from the ${\cal N}=4$ SYM theory: The
  continuous black line describes the flow between the UV fixed point
  and the  IR ${\cal N}=1$ Leigh-Strassler fixed point, induced by
  turning on $m_3 \neq 0$. The blue curve describes an ${\cal
    N}=1$  flow induced by turning on, in addition to $m_3$, operators
  that remain relevant in the IR. 
The red curve corresponds to an ${\cal N}=1$ flow, for some choice of
  relevant couplings,  which just misses the LS fixed point.
}
\label{RGflows}
\end{center}
\end{figure}

Given the flow between two supersymmetric fixed points it is possible
to arrange a 
situation where the theory enters a walking regime. This is achieved
by further deforming the flow by a set of ${\cal N}=1$ 
chiral operators ${\cal O}_i$ (superpotential deformations) which
are relevant at the IR fixed point, with a corresponding set of
couplings ${h_i}$:
\be
{\cal L}_{{\cal N}=4} \,\, \to \,\, {\cal L}_{{\cal N}=4} + \int d^2\theta \, \left(m_3
\Tr\Phi_3^2 + \sum h_i{\cal O}_i\, \right) +\,h.c. 
\ee
We focus on supersymmetric deformations for the usual reasons of
technical convenience, but non-supersymmetric deformations could just
as well be considered. Assuming that the deformed theory has  
at least 
one isolated, supersymmetric ground state ${\cal G}$, one can always
choose the couplings $h_i$ to be {\it parametrically small} so the
flow gets as close as possible to the LS fixed point, before the
relevant operators ${\cal O}_i$ drive the theory to the IR vacuum  
${\cal G}$.\footnote{For supersymmetric deformations, holomorphic
  dependence on the couplings $h_i$ precludes phase transitions as the
  couplings are dialled.} 
The idea is summarized in figure \ref{RGflows}.  

The above scenario is quite general and it is useful to 
have an illustrative example at hand, although we do not
wish to restrict attention solely to it.  Such an example is provided 
by the ${\cal N}=1^*$
deformation of the ${\cal N}=4$ theory, wherein we introduce masses 
$m_i$ for each of the $\Phi_i$. The set of vacua of this theory is
well understood on field-theoretic grounds 
\cite{dw, dorey, doreykum}, and is known to include confining vacua. 
A specific pattern of masses such that $m_1=m _2=m$ and
$m_3=\Lambda_{UV}$, with  
$m \ll \Lambda_{UV}$, realizes the desired walking scenario discussed above.  
Since the flow ends exactly at the LS FP in the case in which
$m=0$, it must pass close to the FP if $m \ll \Lambda_{UV}$. In this case,
the extent of the walking region is controlled by the tunable
parameter $m/\Lambda_{UV}$. This shows that, in principle, 
deformations of ${\cal N}=4$ SYM can furnish  examples of walking
theories in which to pose some of the questions discussed above. 
     
However, it is not {\it a priori} clear which deformations among those
that exhibit walking behaviour also yield a light scalar to be 
identified as a pseudo-dilaton. The answer to this question depends on
the dynamical details, and for this reason the first steps that we
will take in this paper will not be tied to a specific deformation of
the LS flow.  

Changing the coupling constant in ${\cal N}=4$ SYM is an exactly
marginal deformation, so one can choose the UV fixed point to be
weakly or strongly coupled. Regardless of this choice, however, the
theory in the vicinity of the IR fixed point and beyond is inevitably
strongly coupled. While field-theoretic methods may still be useful to
study some aspects of the dynamics in this regime (e.g.~the
calculation of condensates of chiral operators), other aspects require
a different method. These include, for example, the calculation of the
anomalous dimensions of some operators in the vicinity of the LS FP
or the existence of a light dilaton. 

In this paper, we will use the dual string description of the flow
from the ${\cal N}=4$ theory to the Leigh-Strassler SCFT, due to Pilch
and Warner, to address some of these questions. We will focus on
questions that can be answered using only our knowledge of the
Pilch-Warner solution. In particular, we will not address the string
dual description of the relevant deformations that take the theory
away from the LS fixed point.

Our main results are contained in Sections \ref{sec:D7} and
\ref{Sec:3}. In Section \ref{sec:D7} we will determine the anomalous
dimensions of operators of interest in the walking region and compare
them to the corresponding dimensions at the UV FP. We emphasize that
this calculation does not require knowledge of the string dual of the
entire LS flow but only of the LS FP itself. 
For this reason, the results of this calculation will apply to {\it
  any} deformation of the LS flow that one may wish to consider.  
The operators we consider include quark bilinears, which we
describe as fields on probe D7-branes. In Section \ref{Sec:3} we turn to the
existence of a light dilaton in the spectrum. Addressing this question
rigorously does require knowledge of the string dual of the entire
flow (inlcuding the dynamics that drives the theory out of the LS
fixed point), which is not presently available. In order to circumvent
this 
difficulty, we introduce UV and IR cut-offs on the PW solution. The UV
cut-off is a technical convenience and it could be removed. The idea
behind the IR cut-off is that the first part of the flow we are
interested in, represented by the first part of the red curve in
figure \ref{RGflows}, including the walking region, is well
approximated by the LS flow. Cutting-off this solution in the IR is
therefore a crude attempt at modeling, in the spirit of hard-wall models
\cite{hardwall}, the true dynamics that
eventually leads to confinement in the 
deformations of ${\cal N}=4$ SYM of the kind described above. 
Although we find a light
composite state in the spectrum whose mass scales with the size of the
walking region as expected for a light dilaton, we also emphasize that
whether this state persists in the exact theory remains an open
question. In Section \ref{sec:discuss} we discuss possible future directions to
address this question in the exact theory. 

The reader familiar with the necessary background can safely skip to
Section \ref{sec:D7}. However, we have included some additional
sections for the benefit of the unfamiliar reader. In Section
\ref{WTC} we review relevant aspects of WTC. We emphasize the
motivation for the several scales  
$\Lambda_0 \lsim \Lambda_W \lsim \Lambda_I \ll \Lambda_{\ast}$ and for
our focus on the walking region $\Lambda_I<\mu<\Lambda_{\ast}$. In
Section \ref{section2} we review several aspects of the ${\cal N}=4$
theory and its deformations, and we elaborate on the walking scenario
discussed above. In Section \ref{Sec:2} we review the supergravity
description of the LS flow. This section also contains a Wilson-loop
calculation of the quark-antiquark potential along the flow that has
not appeared before.   
Some technical details have been relegated to the Appendices. 

\section{Walking technicolor}
\label{WTC}
In this brief section we remind the reader about why WTC is an
interesting alternative to weakly-coupled models of EWSB. In
particular, we want to summarize the reasons why we are interested in
a theory that has several separated scales ($\Lambda_0 \lsim \Lambda_W
\lsim \Lambda_I \ll \Lambda_{\ast}$), why we would like it to be
quasi-conformal  in the range $\Lambda_I<\mu<\Lambda_{\ast}$ (the
walking region), and why we want large 
anomalous dimensions in this range. 
We dispense with specifying any details of the microscopic realization
of WTC because, since they only affect the physics above the scale
$\Lambda_{\ast}$, they are irrelevant for the arguments presented
here. Much more extensive and detailed discussions can be found
elsewhere~\cite{reviews,P}.  
In reading this section, one should remember that in this paper we are
mostly concerned with the physics taking place in the walking
region. It should also be kept in mind that 
many observables in the context of dynamical EWSB (technicolor) are
hard to calculate using conventional methods because of the associated
strong interactions. For this reason, most of the general arguments we
summarize here are based on order-of-magnitude estimates.

The basic idea of TC is to replace the Higgs sector of the SM by a
completely new, strongly-coupled interaction (called `techni-colour'
after the colour interaction of QCD) together with a corresponding set
of new degrees of freedom (techni-quarks $Q$, techni-mesons $M$,
techni-baryons $B$, etc).  
This strongly-coupled theory will in general have some large internal
global symmetry $G_f$, spontaneously broken by the formation of     
condensates triggered by the strong interaction itself to some
subgroup $G_f\rightarrow H_f$.  
By identifying an $SU(2)\times U(1)\subset G_f$ subgroup of such a
global symmetry  
with the electro-weak gauge symmetry, one can induce EWSB provided
$H_f \cap (SU(2)\times U(1)) = U(1)$, with the latter being the $U(1)$
associated to electromagnetism. 

Let us therefore start from low energies and motivate, on
phenomenological grounds, the requirements we wish to impose on the
strongly coupled theory.  In doing so we will provide a physical
meaning for each of the four dynamical scales in the theory.  
The first condition is that the theory should not contain
experimentally-unobserved light states that couple to the SM. 
The scale $\Lambda_0$ at which the theory confines must therefore
provide a mass gap that is large enough to evade all direct searches
for new (techni-)particles.    

The scale $\Lambda_W$ is the characteristic scale of the condensate
that breaks spontaneously the $SU(2)\times U(1)$  
symmetry. In principle, this might simply be the scale 
$\Lambda_0$ itself, as in QCD, but this need not be the case in
general.
\footnote{The Sakai-Sugimoto model \cite{SS} provides a holographic
  example in which the two scales can be arbitrarily different.} 
The physics of precision parameters, most importantly 
$\hat{S}$~\cite{PT,Barbieri}, is controlled by 
$\Lambda_W/\Lambda_0$ and by model-dependent, hard-to-calculate  coefficients.
Hence one should consider $\Lambda_W$ and $\Lambda_0$ as independent,
though not necessarily parametrically separated, parameters. 

A serious phenomenological problem of TC is that one also has to
implement a mechanism that provides a mass for the SM fermions $f$. In
the absence of a Higgs field, the operator of lowest dimension that
can provide such a mass is of the form 
${\cal O}_6 = \bar{f}f\bar{Q}Q$, 
which on the basis of naive power counting  
is highly irrelevant. This produces two problems. First of all, if one
wants to have a 
UV-complete theory, one has to  
provide an origin for such an operator. This is achieved by embedding
TC into extended technicolor (ETC)~\cite{ETC}, 
the basic idea of which is to unify the family symmetries of
the SM and the gauge symmetry of TC 
into a strongly coupled theory. In this theory, symmetry-breaking yields,
below some scale $\Lambda_{ETC}$, 
an effective theory consisting of TC and a series of higher-order
operators such as ${\cal O}_6$ coupling  TC fermions and SM fermions.

At this point one is left with a second, even more serious
problem, which is the reason why walking  
enters into the game. 
In implementing ETC, one also produces 4-fermion operators of the form
$\bar{f}f\bar{f}f$. These involve SM fermions only, and therefore
contribute directly to FCNC processes. Finding a model of ETC that
avoids inducing excessively-large contributions 
to FCNC processes is tightly bound within the search for models that
explain the large hierarchies between SM fermion masses.  
For the most part this search reduces to a non-trivial exercise
in model building, in which 
one constructs (tumbling) multi-scale  ETC models that implement a
mild version of the Glashow-Iliopoulos-Maiani (GIM) suppression (see
for instance~\cite{APS}).  
What is unavoidable, though, is the (indirectly related) fact that a
new-physics source of 
contributions to the $\hat{T}$ 
precision parameter~\cite{PT,Barbieri} is introduced at
the lightest ETC scale, which contains the dynamical origin of the 
huge top-bottom  mass splitting. Ultimately, this means that one must
at least  require that all ETC scales be 
larger than the  bound deduced via naive dimensional analysis from
$\hat{T}$: $\Lambda_{ETC}\geq\Lambda_{\ast}\gsim 5$ TeV. 

There is hence a scale separation $\Lambda_W \ll \Lambda_{\ast}$. If
naive dimensional analysis  
were to apply, this would mean that the mass of the top quark should
be $m_t\sim \Lambda_W^3/\Lambda_{\ast}^2$, 
and it would be impossible to justify the experimental fact that
actually $m_t\sim {\cal O}(\Lambda_W)$. 
Walking overcomes this difficulty: if the TC theory, for energies
below $\Lambda_{\ast}$, is very close to an approximate fixed point, 
all the operators will inherit their dimensionality not from the naive
counting, but rather from the dimensions of  
 the CFT living at the fixed point. Hence, large anomalous dimensions
 are expected,
affecting all the physical quantities important at long distances,
and the scaling
of the mass of the top would be $m_t\sim
\Lambda_W^3/\Lambda_{\ast}^2\times
(\Lambda_{\ast}/\Lambda_W)^{-\gamma}$, 
with $\gamma$ the anomalous dimension of the chiral condensate.
In this way, large masses for the top can be accommodated, and the
basic phenomenology of the SM is successfully reproduced. 

As stated above, in the deep IR the theory must confine and break
chiral symmetry. 
A conformal theory cannot do so. It must hence be the case that some
physical effect yields
breaking of  scale invariance characterized by a scale $\Lambda_I \ll
\Lambda_{\ast}$, so that the theory is approximately scale invariant  
only in the (walking) region $\Lambda_I<\mu<\Lambda_{\ast}$.
It is usually assumed that $\Lambda_I=\Lambda_W$. Again, it is clear
that there is some relation between  
 $\Lambda_0$, $\Lambda_W$ and $\Lambda_I$, but in general there is no
 obvious reason why they must coincide. 
 Hence one should treat $\Lambda_I$ 
as an independent scale, requiring only that
$\Lambda_0\leq\Lambda_W\leq\Lambda_I$. 

Walking technicolor has also another, less explored but possibly even
more important consequence. 
While ETC has a very large gauge symmetry but a limited amount of
global symmetry, below $\Lambda_{\ast}$ 
the resulting TC model typically has very large global symmetries
(typical QCD-like examples yield at least an $SU(8)_L\times SU(8)_R$ 
symmetry), which are broken only by the higher-order operators of the
form $\bar{Q}Q\bar{Q}Q$ arising at or above the scale
$\Lambda_{\ast}$. 
If there were no large anomalous dimensions, this would immediately
imply the presence of large numbers of pseudo-Nambu-Goldstone bosons
(techni-pions or techni-axions) due to the spontaneous breaking of the
global symmetry (see for 
instance~\cite{AW}). Some of them would carry electro-weak
interactions and have typical masses in the GeV-range, well below the
experimental exclusion bounds, thus invalidating the whole
model. However, because of the same arguments as for the top quark,
the large 
anomalous dimensions could enhance these masses above the experimental
bounds. Furthermore, one might even speculate that the presence of
such higher-order operators might play a role in the very mechanism
leading the theory away from the IR 
 fixed point, and ultimately towards confinement and chiral-symmetry
 breaking.

 To summarize, 
 a generic TC theory potentially  suffers from four phenomenologically
 unacceptable features which, in order of severity, are: (i) A tension
 between the heavy mass of the top and the smallness of the $\hat{T}$
 parameter; (ii) the presence of very light techni-pions; (iii) large
 contributions to precision parameters such as $\hat{S}$; and (iv) a
 potential problem in identifying an ETC model that explains the
 SM-fermion mass hierarchies while 
suppressing FCNC. Avoiding all of these severely restricts the viable
 candidates. WTC is one such candidate characterized by at least the
 four scales $\Lambda_0\leq \Lambda_W\leq \Lambda_I\ll \Lambda_{\ast}$
 and by the presence of large anomalous dimensions in the range
 $\Lambda_I<\mu<\Lambda_{\ast}$. All the ETC physics takes place above
 the scale $\Lambda_{\ast}$ and is of no concern for this
 paper. Rather, our more restricted goal is the study of a model whose
 dynamics in the energy range $\Lambda_I<\mu<\Lambda_{\ast}$ is
 quasi-conformal and strongly-coupled, so that large, non-perturbative
 anomalous dimensions are present. In particular, this means that we
 will not discuss the physics below the scale $\Lambda_I$ either, and
 hence we can (for the time being) ignore completely electro-weak
 symmetry breaking and confinement.   

We close this section by reminding the reader that an active  program
is underway in the context of lattice field theory --- see
e.g.~\cite{lattice} and references therein. The numerical study of a
large class of gauge theories with various field contents aims at (i)
identifying precisely the conditions under which a theory exhibits
conformal or quasi-conformal (walking) behavior in the IR, and (ii)
studying in such models physical observables such as mass spectra,
anomalous dimensions, etc. Proposals such as ours provide a
complementary, analytical approach in which similar questions can be
posed and addressed.

\section{Walking from deformations of ${\cal N}=4$ SYM}
\label{section2}

A natural setting for investigating the walking scenario described
above, within a potentially consistent holographic framework, is
provided by certain relevant deformations of 
${\cal N}=4$ supersymmetric $SU(N)$ gauge theory which 
drive the theory arbitrarily close to a non-trivial 
conformal fixed point at low energies. 
The essential idea is to consider an ${\cal N}=1$ supersymmetric 
deformation of the 
${\cal N}=4$ UV fixed point,
such that deep in the IR the theory confines with a mass gap, whilst
at intermediate energy scales the theory exhibits `almost conformal'
behaviour as it stays in the vicinity of an ${\cal N}=1$ superconformal
fixed point. The ${\cal N}=1$ fixed point in question is the one
discovered by Leigh and Strassler \cite{LS}, which we review
below. For this discussion, it is  
convenient to view the ${\cal N}=4$ theory as consisting of an  ${\cal
N}=1$ vector multiplet ${\cal W}_\alpha$ and three adjoint chiral
multiplets $\Phi_a$, $a=1,2,3$. The superpotential of the
theory is
\begin{equation}
W_{{\cal N}=4}=-{\tau\over 16\pi i }{\rm Tr}\left({\cal W}_\alpha^2\right) 
+\frac{1}{g^2_\mt{YM}}{\rm Tr}\left(\Phi_3[\Phi_1,\Phi_2]\right),
\end{equation}
where %DDD Turned the following into an independent eq %MMM
\be
\tau = \frac{4\pi i}{g^2_\mt{YM}}+\frac{\theta}{2\pi}
\ee
is the
complexified gauge coupling. In this picture, an 
$SU(3)\times U(1)$ subgroup of the full $SU(4)$
R-symmetry of the ${\cal N}=4$ theory is manifest. 

The addition of a  supersymmetric mass term for one of the chiral
multiplets,
\begin{equation}
\Delta W = \frac{1}{2 g^2_\mt{YM}}\,m_3 \Tr \Phi_3^2 \,,
\end{equation}
makes the ${\cal N}=4$ theory flow to
a strongly interacting ${\cal N}=1$ SCFT in the IR
\cite{LS}. Integrating out the massive field for energies below $m_3$
yields the quartic superpotential 
\EQ
{
W= -\frac{\tau}{16\pi i}\Tr{\cal W}_\alpha^2-
\frac{1}{m_3 \,g^2_\mt{YM}}\Tr[\Phi_1,\Phi_2]^2.
\label{quartic}
}
Following the arguments of \cite{LS, ails}, in the deep infrared this
theory  flows to a fixed point wherein 
the anomalous dimension of each of the two light adjoints is
$\gamma_\phi=-1/4$ and the quartic superpotential is actually
an exactly marginal operator --- we define the anomalous dimension of
an operator with canonical dimension $d$ as $\Delta = d+ \gamma_\phi$, 
where $\Delta$ is the scaling
dimension in the IR. The corresponding marginal coupling
inherits the $SL(2,{\mathbb Z})$ duality property of the gauge
coupling of the `parent' ${\cal N}=4$ gauge theory \cite{ails}. 
The presence of the two massless adjoint chiral multiplets implies that
the theory, and in fact the entire flow from ${\cal N}=4$ SYM to the IR
SCFT, has an $SU(2)\times U(1)_R$ global symmetry. The $U(1)_R$ factor
is identified with the non-anomalous R-symmetry of the IR ${\cal N}=1$ 
SCFT.   
In view of \eqn{quartic} the scalar fields $\Phi_{1,2}$ must each
carry R-charge $R=+1/2$ at the fixed point, so that the R-charge of
the superpotential is $R=2$. The ${\cal N}=1$ superconformal algebra
then fixes the scaling dimension of all chiral operators in terms of
their R-charges as  
\be
\Delta= \frac{3}{2} \, |R| \,. 
\label{delta}
\ee
It follows that all
supersymmetry-preserving relevant defomations (F-terms) can also be
identified at the Leigh-Strassler fixed point. 
These include  single-trace operators which are quadratic and cubic in
the fields, such as
$\Tr\Phi_1^2\,,\Tr\Phi_1\Phi_2\,,\Tr\Phi_1^2\Phi_2$, etc. Chiral
operators which are quartic in the scalar fields will be marginal at 
the fixed point, and some of them could be marginally relevant.

The theory below the scale $m_3$ is $SU(N)$ pure ${\cal N}=1$ SYM
coupled to two adjoint chiral multiplets. This theory by itself is
asymptotically free, but in our construction   it matches on to ${\cal
  N}=4$ SYM at the scale $m_3$. Above that scale the presence of the
third chiral multiplet makes the theory conformal. Since the theory
below the scale $m_3$ is asymptotically free, it must become strongly
coupled at a dynamically-generated scale $\Lambda_{\ast} \leq m_3$. In
general, the relationship between these two scales receives (unknown)
instanton contributions and takes the form  
\be
\Lambda_{\ast}^N= m_3^N f(q) \sac \mbox{with } 
q=e^{2\pi i \tau} \,.
\ee
However, if the $\N=4$ 't Hooft coupling is weak, i.e.~if $\gym^2 N
\ll 1$, then $f(q)=q+\cdots$ \cite{ails}. Assuming for simplicity that
the theta-angle vanishes this immediately implies the more familiar
relation 
\EQ
{
\Lambda_{\ast}\sim m_3\;e^{2\pi i \tau/N} 
\sim m_3 e^{-8\pi^2 / \gym^2 N} \,.
}
We expect the theory to flow to the IR CFT only below the
scale $\Lambda_{\ast}$~\cite{ails}.
The existence of the IR fixed point itself is inferred from the exact
beta-functions for the gauge coupling and the coefficient of the
quartic superpotential.  
We may now consider additional perturbations of the theory by operators
which are relevant at the IR fixed point,
\EQ
{
W \to W+ \sum_i h_i {\cal O}_i \,,
}
such that the scales associated to the dimensionful couplings $h_i$
are much smaller than $\Lambda_*$. The theory will then be
quasi-conformal for a large range of energies below $\Lambda_*$, 
until the effect of the relevant couplings $h_i$ becomes important and
drives the theory to a vacuum  with a mass gap (for a suitable choice of
operators ${\cal O}_i$). 

As we have already seen in Section \ref{intro}, 
the simplest example of this situation occurs in the
mass deformation of ${\cal N}=4$ SYM that incorporates
masses $m_1,m_2,m_3$ for all three chiral multiplets, such that
$m_1=m_2=m \ll \Lambda_{\ast}$,
which would leave the theory looking conformal for all energy scales 
$\Lambda_I < \mu < \Lambda_{\ast}$. Here $\Lambda_I \propto m$ with a
constant of proportionality that can, in principle, be a non-trivial
function of the marginal coupling characterizing the ${\cal N}=1$
fixed point. 
The deformation of the ${\cal N}=4$ theory by three  
supersymmetric non-zero mass terms is also known as the ${\cal N}=1^*$
theory. When the scalar VEVs in the field theory vanish, the IR
dynamics of 
the theory is qualitiatively similar to pure ${\cal N}=1$ SYM theory
in that a mass gap and a gluino condensate are   
dynamically generated. For weak UV 't Hooft coupling $g_\mt{YM}^2N\ll 1$,
the strong coupling scale \mbox{$\Lambda_0= (m^2 m_3)^{1/3}\,e^{2\pi
  i\tau/3N}$} will be lower than $\Lambda_I \sim m$, so that the
confining scale is separated from the walking regime.
The vacuum and phase structure, as well as the corresponding
chiral condensates, can all be determined precisely \cite{dw, dorey,
  doreykum, adk, mm} and their functional forms depend in a rather simple way
on the mass parameters $m_i$.  The main point we want to make here is
that the ${\cal N}=1^*$ theory %DDD Changed phrasing
in the limit $m \ll m_3$ provides a realization of the walking
scenario we have described. We will come back to this point in Section
\ref{sec:discuss}. 

We also emphasize, however, that ${\cal  N}=1^*$ should be viewed as
just one example within a wider class of theories that realize the
walking dynamics that we are interested in. 
Other examples include cubic- and perhaps
quartic-superpotential\footnote{In this case we will have to treat the
  theory with an  explicit UV cut-off, since quartic superpotential
  interactions  
will be irrelevant in the ${\cal N}=4$ theory.}
deformations of ${\cal N}=4$ SYM (always accompanied by a large mass
$m_3$) which remain relevant or marginally relevant at the
Leigh-Strassler fixed point. A detailed study of the vacuum structure
and condensates of such deformations would be extremely interesting to
pursue, but is beyond the scope of the present discussion. 

\section{Supergravity description of the Leigh-Strassler flow}
\label{Sec:2}

In this section we first review the basic setup of the dual
supergravity description of the flow from ${\cal N}=4$ SYM to the LS
${\cal N}=1$ SCFT, which is applicable at strong coupling and large
$N$. 
This includes the five-dimensional supergravity sigma-model 
and its uplift to ten dimensions as proposed by Pilch and
Warner~\cite{PW}. Next we discuss in some detail the space of all
possible solutions to the BPS equations of the system, by mapping out
and characterizing the flows in terms of the positions of all the fixed
points and singularities. To some extent, the content of this analysis
is already known in the literature, and can be summarized by saying
that out of all possible solutions to the background equations, only a
one-parameter family is fully acceptable on physical grounds. This
family corresponds to solutions that exactly interpolate between the
${\cal N}=4$ and the LS fixed points. We find it useful  to summarize
these results for the sake of completeness and also to fix
notation. Finally, we examine properties of the flow towards the fixed
points by using Wilson lines to probe the IR geometry.

\subsection{The five-dimensional supergravity sigma-model}

We start from the $SU(2)\times U(1)$-invariant truncation of
five-dimensional  ${\cal N}=8$ supergravity, which contains the ${\cal
  N}=2$ supergravity multiplet. 
In~\cite{PW} it is shown that there exists a further truncation,
obtained by 
requiring invariance with respect to a second $U(1)$, that 
reduces the system to just two scalars, $\chi$ and $\alpha$. These are
dual to a real mass for a fermion and a scalar, respectively, which
together form one of the ${\cal N}=4$ chiral multiplets, $\Phi_3$.  We
study the supergravity sigma-model of these two scalars below.   

As we will see, besides a trivial UV fixed point dual to ${\cal N}=4$
super-Yang-Mills, the equations admit another critical point. The
latter preserves ${\cal N}=2$ supersymmetry in five dimensions
\cite{KPW}, corresponding to an ${\cal N}=1$ SCFT in four dimensions.
\footnote{We remind the reader that the five-dimensional ${\cal N}=2$
  superalgebra has  the same number of supercharges as the
  four-dimensional ${\cal N}= 1$ 
super-\emph{conformal} algebra.} 
The two fixed points are connected by a one-parameter family of
non-trivial solutions dual to the LS flow, 
where the parameter is dual to the supersymmetric mass that specifies
the deformation. 
Four-dimensional ${\cal N}=1$ supersymmetry and $SU(2)\times U(1)$
symmetry are preserved along the entire flow. The $U(1)$ factor
corresponds to the R-symmetry of the four-dimensional ${\cal N}=1$
superalgebra, whereas the $SU(2)$ symmetry is dual to the global
symmetry that rotates the two chiral multiplets into one another. On
the supergravity side both symmetries are realized as isometries of
the solution.  

The five-dimensional truncated supergravity action is given by
\beqs
{\cal S}&=&\int\di^4x\,\di
r\sqrt{-g}\left[\frac{1}{4}R\,-\,\frac{1}{2}G_{ab}g^{MN}\partial_M\phi^a\partial_N\phi^b\,-\,V(\phi)\right]\,, 
\label{5ds}
\eeqs
where $\phi^a=(\chi,\alpha)$.
%and $y^M=(x^{\mu},r)$.
Here, we write the five-dimensional metric as
\beqs
\di s^2_{1,4} &\equiv& e^{2A} \eta_{\mu\nu}\di x^{\mu}\di x^{\nu}\,+\, \di r^2\,,
\label{ds52}
\eeqs
and the sigma-model metric is
\beqs
\,-\,\frac{1}{2}G_{ab}g^{MN}\partial_M\phi^a\partial_N\phi^b&=&-\frac{1}{2}(\partial \chi)^2-3(\partial \alpha)^2\,.
\eeqs

We assume that all the functions defining the background depend only
on the radial direction $r$, and not on the $x^{\mu}$.
It is also 
possible to rewrite the system of equations in terms of a
superpotential $W$, so that the scalar potential is 
\beqs
V&=&\frac{1}{2}G^{ab}W_aW_b\,-\,\frac{4}{3}W^2\,,
\eeqs
where $W_a={\partial W}/{\partial \phi^a}$.
The equations for the background then reduce to
\beqs
\partial_{r} A &=& -\frac{2}{3} W\,,\label{eqforA} \\
\partial_{r} \phi^a&=&G^{ab}W_b\,, \label{eqforphi}
\eeqs
where in the language of~\cite{PW} we set $L=1$ and $g=2/L=2$ to simplify the notation.
The superpotential is given by 
\beqs
W&=&\frac{e^{-2\alpha}}{4}\left[\cosh(2\chi)\left(e^{6\alpha}-2\right)-\left(3e^{6\alpha}+2\right)\right]\,, 
\label{eq:superpot}
\eeqs
and the resulting scalar potential is
\beqs
V&=&
\frac{1}{2} e^{-4 \alpha}
  \cosh^2 \chi \left(-3 - 4 e^{6 \alpha} + \cosh 2 \chi + 
   e^{12 \alpha} \sinh^2 \chi\right)\,.
\eeqs
For completeness we list the general
 set of second-order equations:
\beqs
6A^{\prime\,2}&=&G_{ab}\phi^{\prime\,a}\phi^{\prime\,b}\,-\,2V\,,\\
3A^{\prime\prime}+6 A^{\prime \,2}&=&-G_{ab}\phi^{\prime\,a}\phi^{\prime\,b}\,-\,2V\,,\\
\phi^{a\,\prime\prime}&=&-4A^{\prime}\phi^{a\,\prime}-{\cal
  G}^{a}_{\,\,\,bc}\phi^{b\,\prime}\phi^{c\,\prime}+G^{ab}\frac{\partial
  V}{\partial \phi^b}\,, 
\eeqs
where primed quantities denote derivatives with respect to $r$,
and where the sigma-model connection is trivial:
\beqs
{\cal G}^{a}_{\,\,\,bc}&=&\frac{1}{2}G^{ad}\left(
\frac{\partial G_{db}}{\partial \phi^c}
+\frac{\partial G_{dc}}{\partial \phi^b}
-\frac{\partial G_{bc}}{\partial \phi^d}
\right)\,=\,0\,.
\eeqs

%%%%%%%%%%%%%%%%%%%%%%%%%%%%%%
\subsection{Lift to ten dimensions}

From solutions to the 
five-dimensional truncation one can obtain the full 
ten-dimensional type IIB supergravity solution
by using the lift proposed in~\cite{PW} (see also~\cite{KPW}).
The axion/dilaton system of scalars 
is trivial along the flow, and hence can be ignored for our discussion.
The ten-dimensional metric (in this case there is no difference
between Einstein-frame and string-frame) 
depends on the five-dimensional part $ds^2_{1,4}$ 
(\ref{ds52}) and on the metric of
the internal space,  $\di s_5^2$, which is a deformation of the round $S^5$. 
We  parameterize the internal space by the five angles
 $\theta\in [0,\pi/2]$, $\alpha_1\in [0,\pi]$, $\alpha_2\in [0,2\pi)$,
 $\alpha_3\in [0,4\pi)$ and $\varphi\in [0,2\pi)$. 
 This choice of coordinates makes the symmetries manifest.
The $SU(2)$ symmetry acts on the $\alpha_i$ angles, which will enter the solution
through
the $SU(2)$ left-invariant forms
\bea
\sigma_1\,&=&\, \cos\alpha_3 \di\alpha_1\,+\,\sin\alpha_3\sin\alpha_1
\di\alpha_2\,\,,\\
\sigma_2\,&=&\,\sin\alpha_3 \di\alpha_1\,-\,\cos\alpha_3\sin\alpha_1
\di\alpha_2\,\,,\\
\sigma_3\,&=&\,\di \alpha_3\,+\,\cos\alpha_1 \di \alpha_2\,\,,
\label{explicitsigmas}
\eea
normalized so that $2\di \sigma_i=\epsilon^{ijk}\sigma_j\wedge \sigma_k$.

Following~\cite{PW}, we introduce the following definitions:
\beqs
X_1&=&\cos^2\theta+e^{6\alpha}\sin^2\theta\,,
\\
X_2&=&{\rm sech}\chi \cos^2\theta+e^{6\alpha}\cosh\chi \sin^2\theta\,,
\\
\Omega^2&=&\sqrt{X_1}e^{-\alpha}\cosh\chi\,,
\label{defomega}
\\
e_1&=&e^{-\frac{3}{2}\alpha}(X_1)^{1/4}\cosh^{-1/2}\chi\,\di \theta\,,\\
e_2&=&\frac{1}{2}e^{\frac{3}{2}\alpha}(X_1)^{-1/4}\cosh^{-1/2}\chi\,\cos\theta\,\sigma_1\,,\\
e_3&=&\frac{1}{2}e^{\frac{3}{2}\alpha}(X_1)^{-1/4}\cosh^{-1/2}\chi\,\cos\theta\,\sigma_2\,,\\
e_4&=&\frac{1}{2}e^{\frac{3}{2}\alpha}(X_1)^{1/4}(X_2)^{-1/2}\,\cos\theta\,\sigma_3\,,\\
e_5&=&e^{-\frac{3}{2}\alpha}\frac{(X_2)^{1/2}}{(X_1)^{3/4}}\sin\theta\di\varphi
+\frac{1}{2}\frac{e^{\frac{9}{2}\alpha}\sinh\chi\tanh\chi}{(X_1)^{3/4}(X_2)^{1/2}}\cos^2\theta\sin\theta\,\sigma_3\,.
\eeqs
%\end{widetext}
With all of this in place,
the internal metric is just
\beqs
\di s_5^2&=&\sum_i e_i^2\,,
\eeqs
and the ten-dimensional metric is
\beqs
\di s^2_{10}&=&\Omega^2 \di s^2_{1,4} + \di s_5^2\,.
\label{10dmetric}
\eeqs
We see that the metric of the internal space contains the factor
$e_2^2+e_3^2+e_4^2$, which describes the squashed three-sphere whose
isometry group is $SU(2)\times U(1)$. 

Notice that the five-dimensional metric $ds^2_{1,4}$ now appears with a warp
factor $\Omega^2$ that depends explicitly on the internal coordinate
$\theta$. This is a direct consequence of the fact that 
deforming the ${\cal N}=4$ theory by a mass for one of the adjoint
multiplets not only reduces the global symmetry (isometry of the
internal space) but also lifts two sets of flat directions in the
moduli space of vacua. In practice, this means that
all the warp factors of the dual ten-dimensional background 
(both the internal and the non-compact part of the metric)
depend explicitly on the two coordinates $r$ and $\theta$.

While the dilaton and axion are trivial in the solution
\be
\phi = C_{0} = 0\,,
\label{trivialdil}
\ee
the background includes non-trivial Ramond-Ramond (RR) and Neveu-Schwarz (NS-NS) forms. In order to fix notation for
the following, let us explicitly write down the Bianchi identities and equations of 
motion for the form fields:
\bear
 &&dF_3=dH_3=0\,,\qquad \qquad dF_5=H_3\wedge F_3\,\,,\rc
&&d\left({}^*F_3\right) = -H_3 \wedge F_5\,\,,\qquad
d\left({}^*H_3\right) = F_3 \wedge F_5\,\,,\qquad F_5  = {}^*F_5\,.
\eear
 Two further constraints are needed for (\ref{trivialdil}) to be consistent, namely
\be
F_3\wedge {}^*H_3 = 0 \,\,,\qquad  F_3\wedge {}^*F_3 = H_3\wedge {}^*H_3\,. 
\ee
Here the field strengths are defined in terms of the corresponding
potentials as $H_3\equiv dB_2$ and 
$F_p\equiv dC_{p-1} - C_{p-3}\wedge H_3$.
Then, the non-trivial forms present in the Pilch-Warner solution can be written as
\be
C_2 + i B_2 = e^{-i\,\varphi} (a_1 d\theta +i\,a_2 \sigma_3 +i\, a_3 d\varphi)\wedge (\sigma_1 - i\,\sigma_2)\,\,,
\label{C2B2}
\ee
and 
\be
F_5 = \bar F_5 + {}^*\bar F_5
\label{selfduality}
\ee
with
\be
\bar F_5 = 4\, dx^0 \wedge dx^1 \wedge dx^2 \wedge dx^3 \wedge \left[ (\partial_r w) dr +
(\partial_\theta w) d\theta \right] \,\,.
\label{calF5}
\ee
The coefficients $a_1,a_2,a_3$ and the function $w$ in the expressions
above depend on both $r$ and $\theta$ and are given by
\bear
a_1&=& \frac{1}{2} \tanh \chi \,\cos\theta\,,
\label{a1expr}\\
a_2&=& \frac{1}{4} \frac{e^{6\alpha} \tanh \chi}{X_1} \,\cos^2 \theta\,\sin \theta\,,\label{a2expr}\\
a_3&=& -\frac{1}{2} \frac{\tanh \chi}{X_1} \,\cos^2 \theta\,\sin \theta\,,\label{a3expr}\\
w&=& \frac18 e^{4A-2\alpha}
\left(-2 \cosh^2(\chi)  \cos^2\theta + e^{6\alpha} (\cosh(2\chi)-3) \sin^2 \theta 
\right)\,.
\label{wexpression}
\eear
The isometry of the internal part of the metric is $SU(2)\times
U(1)^2$, where the $U(1)$'s are associated with shifts in $\varphi$
and $\alpha_3$. 
However, the isometry of the full background is only $SU(2)\times
U(1)$ because the two-forms (\ref{C2B2}) are only invariant under
shifts that leave the combination $\varphi - \alpha_3$ invariant. 

%%%%%%%%%%%%%%%%%%%%%%%%%%%%%%%%%%%
\subsection{Fixed points and flows}
\label{sec:fp}
We can write explicitly the BPS  equations for the scalars 
\beqs
\partial_{r} \chi&=&
\frac{1}{2}e^{-2\alpha}\left(-2 + e^{6 \alpha}\right)\sinh 2\chi
\,,\rc
\partial_{r} \alpha&=&
\frac{1}{6}e^{-2\alpha}\left[1 + e^{6 \alpha} \left(-3 + \cosh 2 \chi\right) + \cosh 2 \chi\right]
\,,
\label{chialphaeqs}
\eeqs
and start the analysis looking for critical points of the flow.
These include the trivial solution
\beqs
\label{Eq:UV}
\chi\,=\,0\,,&&\,\alpha\,=\,0\,,
\eeqs
to which we will refer as $F_U$ in the following, and the five-dimensional ${\cal N}=2$ fixed points 
\beqs
\label{Eq:IR}
\chi\,=\,\pm\,{\rm arcosh}\frac{2}{\sqrt{3}}\,=\,\pm\frac{\ln 3}{2}\,,&&\,\alpha\,=\,\frac{\ln 2}{6}\,,
\eeqs
to which we will refer as $F_I$ and $\bar{F}_I$. The are no other constant solutions to the BPS equations.

\begin{figure}[htpb]
\begin{center}
\begin{picture}(300,240)
\put(0,0){\includegraphics[width=13cm]{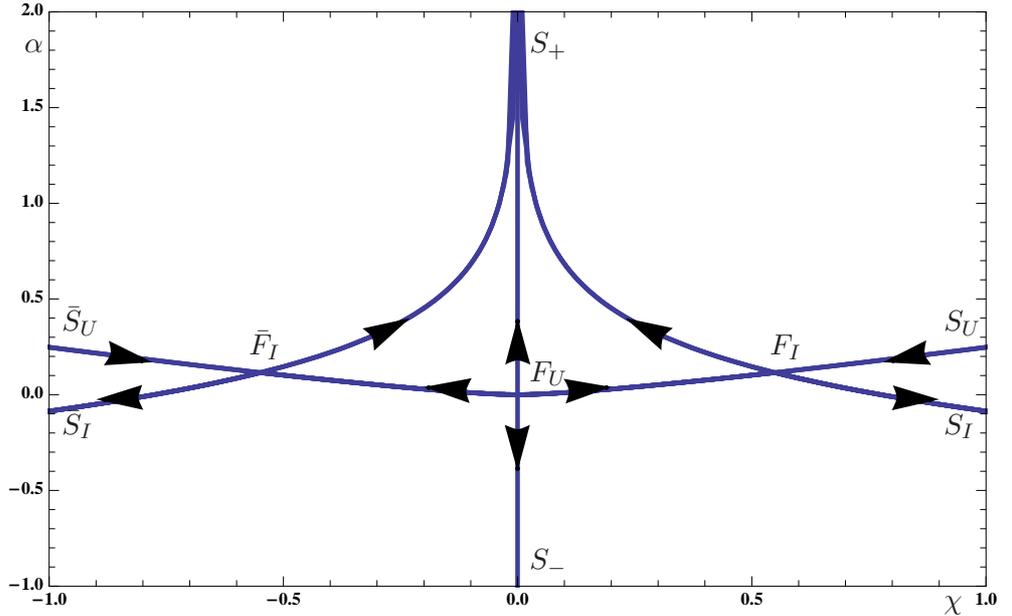}}
\put(350,0){$\chi$}
\put(6,210){$\alpha$}
\put(195,86){$F_U$}
\put(195,210){$S_+$}
\put(195,15){$S_-$}
\put(285,96){$F_I$}
\put(90,96){$\bar{F}_I$}
\put(350,105){$S_U$}
\put(350,66){$S_I$}
\put(20,105){$\bar{S}_U$}
\put(20,66){$\bar{S}_I$}
\end{picture} 
\caption{The $\alpha - \chi$ plane and the flows. Explicitly shown are the fixed points $F_U$, $F_I$ and $\bar{F}_I$,
together with the singularities $S_{\pm}$, $S_U$, $S_I$, $\bar{S}_U$ and $\bar{S}_I$, and the RG lines connecting them.
The arrows indicate the direction from the UV to the IR (decreasing $r$).}
\label{Fig:RG}
\end{center}
\end{figure}

In Fig.~\ref{Fig:RG} we plot the %DDD Changed $\alpha(\chi)$ to $\alpha - \chi$ MMM
$\alpha - \chi$ space, and the various
flows from and to the three fixed points. 
The system is completely symmetric under $\chi\rightarrow -\chi$.
Notice that, besides the fixed points, there are six possible
endpoints for the flows.  For $\chi=0$, the flow out of the fixed
point at the origin goes to $\alpha \rightarrow \pm \infty$ (the
points $S_{\pm}$ in the plot). 
The flow out of the IR fixed points along the relevant deformation can
either end  
towards $\chi\rightarrow 0$ and $\alpha\rightarrow +\infty$ (point $S_+$), 
or towards $\chi\rightarrow +\infty$ and $\alpha\rightarrow -\infty$
(singularity at $S_I$), and similarly for the flows out of
$\bar{F}_I$. 
Finally, flows that end into the IR fixed point $F_I$ could originate
in the UV from $\chi\rightarrow +\infty$ 
and $\alpha\rightarrow +\infty$ (the point $S_U$), and similarly for
the $\chi\rightarrow -\chi$ case. 

%%%%%%%%%%%%%%%%%
\subsubsection{UV fixed point}
Starting from Eqn.~(\ref{Eq:UV}), and making the replacement 
$X_1=X_2=1$ for the UV fixed point, the internal space is simply an $S^5$:
\beqs
\di s_5^2&=&\di \theta^2+\frac{1}{4}\cos^2\theta \left(\sigma_1^2 + \sigma_2^2 + \sigma_3^2 \right)
+\sin^2\theta\di \varphi^2\,.
\eeqs
The superpotential at this minimum is
$
W = -3/2 \,,
$
and hence (setting an integration constant to one)
$
A(r) = r\,.
$
The warp factor is
$
\Omega^2 = 1\,,
$
and  consequently the metric reduces to that of $AdS_5\times S^5$
with radii $L=1$ for both factors:
\beqs
\di s^2_{10}&=&e^{2 r}\di x_{1,3}^2\,+\,\di  r^2\,+\,\di s_5^2\,.
\eeqs
Expanding the scalar potential $V$ around this minimum and normalizing
the $\alpha$-fluctuation canonically through
$\bar{\alpha}=\sqrt{6}\alpha$, 
the mass matrix  in the $(\chi,\bar{\alpha})$ basis is
\beqs
L^2 M_\mt{scalar}^2&=&\left(\begin{array}{cc}
-3&\cr
&-4\end{array}\right)\,.
\label{scalar}
\eeqs
As usual, the dimension of the dual operators are given by the largest
root of the equation $\Delta (\Delta-4)=L^2 M_\mt{scalar}^2$. This is
$\Delta_\chi=3$ for the $\chi$ field, confirming that 
it is dual to a fermion-mass operator. Its most general behaviour near the boundary is
\be
\chi \approx A \, e^{-(4-\Delta_\chi)r} + B \, e^{-\Delta_\chi r} \,,
\label{deltacomp}
\ee
where $A$ and $B$ correspond to the coefficient (the mass) and the VEV
of the operator, respectively. The requirement that the flow be
supersymmetric, however, selects $B=0$. This can be seen by expanding
the superpotential around the same fixed point, 
\beqs
W&=&-\frac{3}{2}\,-\bar{\alpha}^2\,-\frac{1}{2}\chi^2\,+\cdots\,,
\eeqs
and the using the BPS equation \eqn{eqforphi}, which immediately
implies $\chi \sim e^{-r}$. A similar analysis for the $\alpha$ scalar
shows that it is dual to a scalar-mass operator of dimension
$\Delta_\alpha =2$ and that no VEV is present. Thus we conclude that
both directions correspond to the insertion of relevant deformations
(masses), as illustrated in Fig.~\ref{Fig:RG} 
by the instability of $F_U$.

%%%%%%%%%%%%%%%%%
\subsubsection{IR fixed point}

At the IR fixed point one has
\beq
X_1=1+\sin^2\theta
\,, \qquad
X_2=\frac{3+5\sin^2\theta}{2\sqrt{3}}
\,,\qquad
\Omega^2=\frac{2^{5/6}}{\sqrt{3}} \sqrt{1+\sin^2\theta}\,,
\eeq
which yields
\beqs
e_1&=&\frac{3^{\frac14}}{2^{\frac34}} \sqrt[4]{1+\sin^2 \theta}\,\di \theta\,,\qquad
e_2=\frac{3^{\frac14}\cos \theta }{2^{\frac54}
   \sqrt[4]{1+\sin^2 \theta}} \, \sigma_1\,,\\
   e_3&=&\frac{3^{\frac14}\cos \theta }{2^{\frac54}
   \sqrt[4]{1+\sin^2 \theta}}\, \sigma_2\,,\qquad
   e_4=\frac{3^{\frac14}\cos \theta \,
   \sqrt[4]{1+\sin^2 \theta}}{2^{\frac14}\sqrt{3+5\sin^2 \theta}} \, \sigma_3\,,\\
   e_5&=&\frac{ \sin \theta \, \sqrt{3+5\sin^2 \theta}}
   {2^{\frac34}3^{\frac14}(1+\sin^2 \theta)^{3/4}}\, \di \varphi\,+
   \,\frac{ \cos ^2 \theta \,  \sin \theta }{\sqrt[4]{24}
   \sqrt{3+5\sin^2 \theta} \, (1+\sin^2 \theta)^{3/4}}\, \sigma_3\,.
\eeqs
At the IR fixed point the AdS warp factor is given by 
\beqs
A(r)&=&\frac{2^{5/3}}{3} \, r\,,
\eeqs
which numerically means $L_{IR}=3\times 2^{-5/3} L \simeq 0.95 L$, in
agreement with the fact that the curvature in the IR 
should be larger, and with the famous relation
\beqs
\frac{L_{UV}^3}{L_{IR}^3}&=&\frac{32}{27}\,.
\eeqs
We have explicitly written $L_{UV}$ in order to make clear the
comparison between the UV and the IR  fixed points, but recall that we
are taking $L_{UV}\equiv L = 1$ in most of the equations.  

The IR metric is thus
\beqs
\di
s^2_{10\,IR}&=&\frac{\sqrt{1+\sin^2\theta}}{\sqrt{L_{IR}}}\left(e^{2r/L_{IR}}\di
  x_{1,3}^2\,+\,\di r^2\right)\,+\,\di S_{5\,IR}^2\,, 
\label{IRmetric}
\eeqs
which shows the explicit dependence of the metric on the internal
coordinate $\theta$.  We do not write the explicit expression for the forms, 
which are readily  found by inserting (\ref{Eq:IR}) into
 (\ref{C2B2})-(\ref{wexpression}). The IR geometry only has an
 $SU(2)\times U(1)$ isometry as expected for the Leigh-Strassler
 CFT. The $\theta$-dependence of the warp factor is directly related
 to the fact that the $SU(N)$ ${\cal N}=4$ theory has a $6N$-dimensional
 moduli space of vacua whilst the Leigh-Strassler CFT has two massless adjoint
 chiral multiplets and an associated $4N$-dimensional moduli
 space. The moduli space of the IR theory can be  revealed by a probe
 D3-brane immersed in the geometry above (see e.g.~
 \cite{Johnson:2000ic}), which explores the Coulomb branch of the
 moduli space associated to the Higgsing  $SU(N)\to SU(N-1)\times
 U(1)$. The probe brane analysis shows that the supersymmetric vacua
 of the theory are at $\theta=0$, at which point the probe potential
 vanishes and the D3-brane sees a
 four-dimensional moduli space.

Expanding the scalar potential $V$ around the IR fixed-point, the mass
matrix  in the $(\chi,\bar{\alpha})$ basis is 
\beqs  
L_{IR}^2 M_\mt{scalar}^2 &=
\left(\begin{array}{cc}
6&2\sqrt{6}\cr
2\sqrt{6}&2
\end{array}
\right)\,.
\eeqs
%DDDD New text below
The eigenvalues are $\lambda_\mt{irr} = 2(2 + \sqrt{7})$ and
$\lambda_\mt{rel} = 2(2 - \sqrt{7})$, corresponding to a dual
irrelevant operator ${\cal O}_\mt{irr}$ and a dual relevant operator
${\cal O}_\mt{rel}$ of dimensions  
\be
\Delta_\mt{irr} = 3 + \sqrt{7} \simeq 5.65 \sac 
\Delta_\mt{rel} = 1 + \sqrt{7} \simeq 3.65 \,.
\ee
To see the implications of supersymmetry we follow the previous
subsection and expand the superpotential around the IR fixed point,
with the result 
\beqs
W L_{IR}&=&-\frac{3}{2}\,-\delta
\bar{\alpha}^2\,+\sqrt{6}\delta\chi\delta \bar{\alpha}\,+\cdots\,,
\label{deltair}
\eeqs
where $\delta\bar{\alpha}$ and $\delta \chi$ are infinitesimal
deviations from the fixed point. After diagonalization of the BPS
equations  \eqn{eqforphi} we find that the two independent
fluctuations behave near the boundary as  
$X_\pm \approx e^{- r \Delta_\pm}$ with 
\be
\Delta_+ = 4- \Delta_\mt{irr} = 1- \sqrt{7} \sac 
\Delta_-= \Delta_\mt{rel} = 1- \sqrt{7} \,.
\label{deltapm}
\ee
Comparing with \eqn{deltacomp} we see that supersymmetry allows for
the coupling of ${\cal O}_\mt{irr}$ and the VEV of  ${\cal
  O}_\mt{rel}$ to be present, but forces the VEV of ${\cal
  O}_\mt{irr}$ and the coupling of ${\cal O}_\mt{rel}$ to vanish.  
We can infer from this that when the same model is studied with a
finite UV cut-off, as in Section \ref{Sec:3}, the resulting change in
boundary conditions will induce an irrelevant coupling, namely a
double-trace deformation involving  ${\cal O}_\mt{rel}$ with scaling
dimension  
$\Delta^{\prime}= 2(1+\sqrt{7})$.

\subsection{Wilson loops}
\label{subsec:Wilson}

So far we have 
reviewed the construction of the supergravity solution which describes the
flow from ${\cal N}=4$ SYM to the Leigh-Strassler IR fixed point. In
order to probe how the physics changes along the flow, we now discuss
one simple observable that can be computed on the gravity side, namely
the 
expectation value of a Wilson loop describing the interquark potential
for a test quark-antiquark pair. Following standard techniques 
\cite{wilson1,wilson2}, this amounts to computing the energy of a
string with endpoints at the boundary of AdS space that minimizes the
Nambu-Goto action. We will consistently restrict ourselves to string
solutions that lie 
at $\theta=0$, since the geometry at $\theta=0$ correctly describes the
vacuum of the gauge theory in the IR. 
Indeed, if we think of separating a D3-brane and letting the string
hang from it, since the D3-brane can  
only be placed at $\theta=0$ (coinciding with the moduli space of the
field theory),  
the string will remain at $\theta=0$. Excursions away
from $\theta=0$ will necessarily cost additional
energy. %DDD Changed phrasing of what used to be a footnote MMM
Even if the string endpoints are placed at some $\theta\neq 0$, for
large interquark separation the dominant contribution will come from a
long piece of string that will lie very close to $\theta=0$. Therefore
we will not consider $\theta \neq 0$ configurations further. 

\begin{figure}[h]
\begin{center}
\includegraphics[width=10cm]{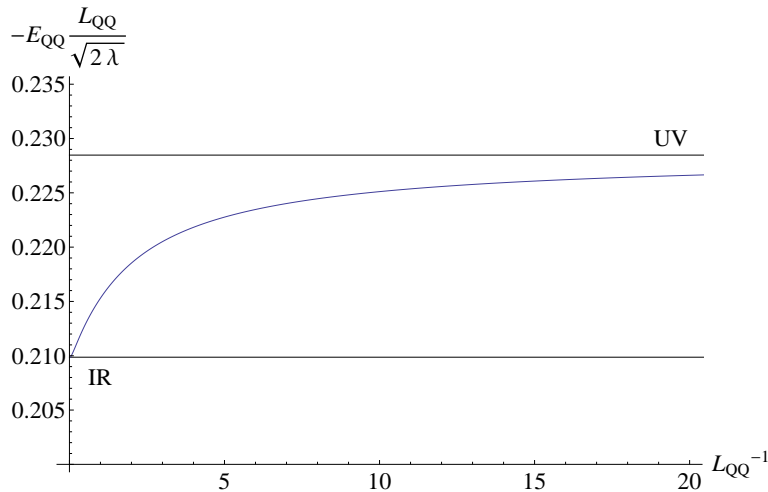}
\end{center}
\caption{Numerical results for the energy and length of the string probing the flow from
${\cal N}=4$ SYM down to the LS fixed point.}
\label{fig:wilsonflow}
\end{figure}

As usual, there is a one-parameter family of solutions depending on
the value $\hat{r}_0$ down 
to which the string descends. Inserting the metric (\ref{defomega}),
(\ref{10dmetric})
 into the general expressions of \cite{wilson2}, one finds
the length and energy of the classical string configuration. The
formally-divergent energy is renormalized by subtracting off the
energy of two infinite straight strings corresponding to the
(infinite) masses of the test quark-antiquark pair. The renormalized 
inter-quark separation and potential energy are then
\bear
L_{\bar Q Q} (\hat{r}_0)&=&  2 \int_{\hat{r}_0}^\infty
e^{-A} \left(e^{-2\alpha + 2\alpha_0}\frac{\cosh^2 \chi}{\cosh^2
    \chi_0}e^{4A-4A_0}-1 
 \right)^{-\frac12} dr \,,\rc
E_{\bar Q Q} (\hat{r}_0) &=& \frac{1}{2\pi \alpha'}2
\left[ \int_{\hat{r}_0}^\infty  
e^{-\alpha}\cosh \chi e^{A}
\left( \left(1- e^{2\alpha - 2\alpha_0}\frac{\cosh^2 \chi_0}{\cosh^2
      \chi}e^{4A_0-4A} \right)^{-\frac12} -1 \right) 
dr \right.
\rc
&&\left. - \int_{-\infty}^{\hat{r}_0}   e^{-\alpha}\cosh \chi e^{A} dr
\right]\,\,.
\eear
The constants $A_0, \alpha_0$ and $\chi_0$ are the values of the
respective functions at $r=\hat{r}_0$.
By numerically integrating these expressions, we can find the binding
energy in terms of the  
quark-antiquark separation, which we plot in Figure (\ref{fig:wilsonflow}). 
From the
plot, we see how the potential interpolates between its UV value at
small inter-quark separation and the IR value at very large
separation, with 
\be
E_{\bar Q Q}^{UV} = - \frac{4\pi^2}{\Gamma\left(\frac14\right)^4}\,\frac{\sqrt{2\lambda}}{L_{\bar Q Q}}\,\,,\qquad
\qquad
E_{\bar Q Q}^{IR} = - \sqrt{\frac{27}{32}}
\frac{4\pi^2}{\Gamma\left(\frac14\right)^4}\,\frac{\sqrt{2\lambda}}{L_{\bar Q Q}}\,\,,
\ee
where we have introduced the 't Hooft coupling of the UV theory 
$\sqrt{2\lambda} =L_{UV}^2 / \alpha'$.  At the IR and UV fixed points
the potential is Coulombic as dictated by conformal invariance.  
The numerical factors in front of the UV and IR expressions for the
potential differ by the quotient of central charges
$\sqrt{c_{IR}/c_{UV}} = \sqrt{27/32}$ at the two fixed points.  

The Wilson loops can also be used to probe the flows that approach the
singular end-points, labelled $S_\pm$ and $S_I$ in
Fig.~\ref{Fig:RG}. As expected, explicit computation of these physical
obervables shows (see appendix \ref{app:sing}) that we should not ascribe
physical meaning to these singular endpoints.

\section{Anomalous dimensions of quark bilinears in the walking region}
\label{sec:D7}

In this section we will introduce flavour fields into the gauge theory
dual to the Pilch-Warner flow. The resulting theory is $SU(N)$ ${\cal
  N}=4$ SYM coupled to $N_f$ ${\cal N}=2$ hypermultiplets in the
fundamental representation of the gauge group. The addition of these
fields breaks the ${\cal N}=4$ supersymmetry at least down to ${\cal
  N}=2$. Although the hypermultiplets contain both bosonic (squark)
and fermionic (quark) fields, in an abuse of language we will
collectively refer to them as `quarks'.  As usual in the `t Hooft
limit, we consider $N \to \infty$ with $N_f$ fixed, so that $N_f \ll
N$.  
As is well known, in this limit the quarks behave as `probes' of the
gluon-plus-matter-dominated dynamics. For this reason they do not
break the conformal invariance of the UV and IR fixed points at
leading order. In other words, their contribution to the beta-function
of the theory is suppressed by $N_f/N$ relative to the order-$N^2$
contribution from the adjoints fields. Correspondingly, in the string
description the quarks  are described by $N_f$  probe D7-branes in the
Pilch-Warner geometry that do not backreact on the geometry at leading
order. The D7-branes  wrap a three-cycle in the compact part of the
geometry. The choice of this three-cycle is dual to the choice of
couplings between the flavour fields and the adjoint fields in the
gauge theory. Depending on the orientation of this three-cycle, these
couplings will preserve or break all or part of the
superymmetries. Here we will focus on a supersymmetric embedding such
that the maximal possible amount of supersymmetry is preserved: ${\cal
  N}=2$ in the UV and ${\cal N}=1$ all along the flow, including at
the IR fixed point. Thus at this point we obtain the large-$N$
Leigh-Strassler SCFT coupled to supersymmetric matter.  

As explained above, our goal is to compute the dimensions at the IR
fixed point of a number of operators built from quark fields. For this
reason, we will focus on the dynamics of the D7-branes in the geometry
dual to the LS fixed point, as opposed to the geometry dual to the
entire flow. For the UV fixed point, this computation was done in
\cite{Aharony:1998xz, Kruczenski:2003be}. Comparing with those
references we will be able to check how dimensions change due to the
RG flow to the Leigh-Strassler CFT. In other words, we will be able to
compute the \emph{anomalous} dimensions of the operators at the IR
fixed point with respect to the UV fixed point. As emphasized in
Section \ref{intro}, the anomalous dimensions that we will obtain will
be valid for \emph{any} deformation of the ${\cal N}=4$ SYM theory
that `walks' near the LS fixed point. For a large class of operators
the anomalous dimensions turn out to be negative and, as expected from
the fact that the fixed point is strongly coupled, large. As reviewed
in Section \ref{WTC}, large negative anomalous dimensions are an
essential ingredient in phenomenologically interesting theories that
exhibit walking behaviour. We will not address the departure of the
theory from the walking region, i.e.~away from the vicinity of the IR
fixed point. We will return to this issue in Section
\ref{sec:discuss}, where we will discuss the possibility of eventually
deviating from the walking region by deforming the D7-branes
embedding.

As mentioned above, since $N_f \ll N$ we do not need to consider the
backreaction of the D7-branes on the background 
geometry \cite{Karch:2002sh} (see \cite{Nunez:2010sf} for a recent
review of holographic models in the Veneziano limit, where $N_f/N$ is
kept fixed).  
The operators we will discuss are bilinears in the quark fields associated to fluctuations of
the D7-branes.  There are two primary reasons for focussing on these 
excitations rather than those  
of the supergravity fields (which are dual to the adjoint matter).  First,
this sector is simpler due to the relatively smaller number of 
degrees of freedom on the D7-branes compared to the number of
supergravity modes. In fact, we will be  able to find the scaling
dimensions of all operators associated to the excitation modes of one
D7-brane.
\footnote{
We will explicitly discuss the bosonic operators, but the dimensions
of their fermionic superpartners follow from supersymmetry, as in
\cite{Kruczenski:2003be}.} 
Second, quark fields in the form of D7-brane probes can
be expected to exhibit interesting behaviour either when the background
geometry is deformed (e.g.~via supersymetry-breaking deformations) or
when the embedding itself is deformed away from the supersymmetric
one. 
In such a stituation, it is possible for the D7-branes 
to bend and display chiral symmetry breaking as shown 
in \cite{Babington:2003vm,Kruczenski:2003uq} %\cite{Kruczenski:2003uq}
 and subsequent papers. Therefore, set-ups with D7-branes can be
 interesting for modelling walking 
 theories as we will discuss further in Section \ref{sec:discuss}.

From now on we will focus on D7-branes sitting at 
$\theta=0$. It can be shown via a kappa-symmetry analysis that these
branes are supersymmetric --- the Killing spinors of the solution 
are given in \cite{Pilch:2004yg}. At $\theta=0$, the probe branes wrap
the squashed three-sphere with metric $e_2^2+e_3^2+e_4^2$ and
$SU(2)\times U(1)$ isometry. From these symmetry considerations it is 
clear that the flavour D7-branes have the effect of introducing
superpotential interactions between $2N_f$ fundamental chiral 
multiplets $(Q^i,\tilde Q_i)$ of the form
\be
W= W_{{\cal N}=4} + \frac{1}{g^{2}_\mt{YM}}\, \left[m_3 \Tr\Phi_3^2+
\sum_{i=1}^{N_f} Q ^i \Phi_3 \tilde Q_i
\right]\,.
\label{compare}
\ee
In order to avoid confusion, below we will denote as $q^i$ the
fermionic component fields of the chiral superfields $Q^i$. The matter
content and the couplings follow from ${\cal N}=2$ supersymmetry in
the UV.  
Upon integrating out $\Phi_3$ we obtain a quartic superpotential that
includes terms of the schematic form, $Q \tilde Q\,Q \tilde Q$.
It follows that, at the Leigh-Strassler IR fixed point, we can use the
non-anomalous (at large $N$) $U(1)_R$ to assign R-charges to the
fundamental scalars (the squarks) $R[Q]=R[\tilde Q]=1/2$.
Eqn.~\eqn{delta} then implies that all the quartic terms in the
superpotential have dimension 3 and are therefore marginal. 

\subsection{Probe D7-brane dynamics}
In the rest of this section we will analyze linear oscillations of a
single D7-brane around the $\theta=0$ solution in the geometry dual to
the LS FP, and compute the dimensions of the associated operators.  
At leading order in $N$, the generalization to multiple overlapping
D7-branes is straightforward \cite{Kruczenski:2003uq}.

The worldvolume action
for the bosonic degrees of freedom 
 of a probe D7-brane is given by the sum of Dirac-Born-Infeld (DBI)
 and Wess-Zumino (WZ) terms:\footnote{Different conventions for the WZ
   term can be found in the literature. Our convention  (\ref{DBIWZ})
   is consistent with  the definition of the RR forms that we have
   adopted.} 
\be
S_{D7}= - T_7 \left( \int d^8x \, e^{-\phi} \sqrt{-\det(P[g] + {\cal F})} 
+ \int P[C] \ e^{-{\cal F}} \right)
\label{DBIWZ}
\ee
where $P[\ldots]$ refers to the pull-back onto the brane.
From the last term of the action, one only selects the
contributions from 8-forms, namely
\be
\int P[C] \ e^{-{\cal F}} \equiv \int \left[
P[C_8] - P[C_6] \wedge {\cal F} + \frac12 P[C_4] \wedge {\cal F}
\wedge {\cal F}+ \cdots \right] \,. 
\ee
The two-form ${\cal F}$ is defined as:
\be
{\cal F} = P[B_2] + 2\pi \alpha' F \,,
\ee
where $F$ is the Abelian field strength of the worldvolume gauge field.
The definition of the pull-back of the metric is:
\be
P[g]_{MN}=\partial_M X^A \partial_N X^B g_{AB} \,,
\ee
where $M,N=1, \dots 8$ are worldvolume indices on the D7-brane, 
while $A,B=1,\dots,10$ are space-time indices,
and the $X^A$ are space-time coordinates. The pull-backs of the
forms are defined in a similar 
fashion.

As mentioned above, we want to analyze oscillations around a classical solution
given by $\theta=0$, with vanishing worldvolume gauge field.
A technical  issue is that around $\theta=0$, the $\theta,\varphi$
coordinate system  
is singular. We make the following change of coordinates to a
Cartesian-like system, which is well-defined in the neighbourhood of 
$\theta=0$:
\be
z_1 = \theta \sin \varphi\,\,,\qquad\qquad z_2 = \theta \cos \varphi \,.
\ee
Since we only need to keep terms in the action which are quadratic in
the fluctuations, we expand the metric up to
second order in the $z$-coordinates and $B_2$ and $C_6$ to first
order in $z_{1,2}$. 
We also find it convenient to 
define
\be
y=L_{IR}^{-\frac14} \,\exp\left(\frac{r}{L_{IR}}\right)\,\,.
\ee
In terms of this the ten-dimensional metric \eqn{IRmetric} dual to the LS fixed point, expanded to quadratic order 
in $z_i$, reads
\bear
ds_{10}^2&=&(1+\frac12 z_1^2 + \frac12 z_2^2)\left(y^2 dx_{1,3}^2 +
  L_{IR}^\frac32\, \frac{dy^2}{y^2} 
\right) 
+\frac{\sqrt3}{4\sqrt2}\Bigg[(1-\frac32 z_1^2 - \frac32 z_2^2)(\sigma_1^2 + \sigma_2^2)+\rc
&+& \frac43(1-2z_1^2-2z_2^2)\sigma_3^2 + 2 dz_1^2 + 2 dz_2^2 + \frac43(z_2 dz_1 - z_1 dz_2)\sigma_3
\Bigg]\,.
\eear
Defining $(dx)^4\equiv dx_0 \wedge dx_1 \wedge dx_2 \wedge dx_3$, 
 to first order in $z_i$'s, the potentials for the form fields are
\bear
B_2 &=& -\frac{1}{4}\left[  z_2 \sigma_1 \wedge \sigma_3 -  z_1 \sigma_2 \wedge \sigma_3
+dz_2 \wedge \sigma_2 +dz_1 \wedge \sigma_1\right]
\,\,,\rc
C_6&=&\frac{y^4}{16}(dx)^4 \wedge \left[ 5 z_2 \sigma_1 \wedge \sigma_3 - 5 z_1 \sigma_2 \wedge \sigma_3
+3 dz_2 \wedge \sigma_2 +3 dz_1 \wedge \sigma_1\right]\,\,,\rc
C_4 &=& - y^4 (dx)^4 \,\,. 
\label{someforms}
\eear
Here $C_6$ is defined by the relation $F_7=-{}^*F_3=dC_6 - C_4 \wedge H_3$.
Notice that $C_2$ does not explicitly enter the quadratic action since
$C_2 \wedge B_2 =0$. 
The four-form
$C_4$ has an extra component with legs along the angles which we have
not written since it 
does not couple to the worldvolume of the D7-brane and does not play any
role in the following. 
It is convenient to perform a gauge transformation and to add a total
derivative to $B_2$, in order  
to define a new, simpler NS-NS potential, namely:
\be
B_2 = -\frac{1}{2}\left[  z_2 \sigma_1 \wedge \sigma_3 -  z_1 \sigma_2
  \wedge \sigma_3\right] 
\,\,.
\label{newB2}
\ee
It is now a lengthy but straightforward computation to find the
quadratic action for the fluctuations. We first 
define the determinant of the worldvolume metric at zeroth order in the
fluctuations,
\be
\sqrt {-g_0}=  \frac{3\sqrt3}{32} y^3  \sqrt{\tilde g_3} \,,
\label{sqrtg0}
\ee
where $\tilde g_3$ denotes the metric on the squashed $S^3$,
\be
d\tilde s_3^2 = \sigma_1^2 + \sigma_2^2 + \frac43 \sigma_3^2 = 
d\alpha_1^2 + \sin^2 \alpha_1 d\alpha_2^2 + \frac43 (d\alpha_3 + \cos \alpha_1 d \alpha_2)^2\,.
\label{squsS3}
\ee
From now on we will denote the squashed three-spehere as $\tilde S^3$. %MMM
The metric (\ref{squsS3})
is of the type discussed in Appendix \ref{app:harmonics} with $a^2=1$, 
$b^2= 4/3$. Introducing the dreibein for the squashed sphere as:
\be
\tilde e^1 = \sigma^1 \,\,, \qquad 
\tilde e^2 = \sigma^2 \,\,, \qquad 
\tilde e^3 = \frac{2}{\sqrt3}\sigma^3 \,\,, \qquad 
\ee
or in components $\tilde e^a = \tilde e_i^a d\alpha^i$, and the
inverse  dreibein as $d\alpha^i = e_a^i \tilde e^a$ such that $\tilde
e_i^a e_a^j = \delta_i^j$, we have
\be
\tilde e= \left(
\begin{array}[h]{c c c}
        \cos\alpha_3  & \sin\alpha_3 & 0\\ 
        \sin\alpha_1 \sin\alpha_3   & -\sin\alpha_1 \cos\alpha_3  &
        \frac{2}{\sqrt3}\cos\alpha_3\\
        0 & 0 & \frac{2}{\sqrt3}
\end{array}\right)
\,,\quad
e= \left(
\begin{array}[h]{c c c}
        \cos\alpha_3  & \frac{\sin\alpha_3}{\sin\alpha_1} & -\frac{\sin\alpha_3\cos\alpha_1}{\sin\alpha_1}\\ 
        \sin\alpha_3   & -\frac{\cos\alpha_3}{\sin\alpha_1} &
        \frac{\cos\alpha_3\cos\alpha_1}{\sin\alpha_1}\\
        0 & 0 & \frac{\sqrt3}{2}
\end{array}\right) \,.
\label{matrixviel}
\ee
Notice that $\sum_a e_a^i e_a^j = \tilde g^{ij}_3$, the inverse metric on the 
squashed three-sphere.
We can now write the full quadratic worldvolume action which,
after subtracting a total derivative, reads:
\bear
-T_7^{-1} S_{D7}= \int d^8x \sqrt{-g_0}\Bigg(-\frac14(z_1^2 + z_2^2)
+
\frac{\sqrt3}{4\sqrt2} g^{MN}(\partial_M z_1 \partial_N z_1 
+ \partial_M z_2 \partial_N z_2  ) +\rc 
+ \frac{1}{2} (z_2  \partial_{\alpha_3} z_1 - 
z_1 \partial_{\alpha_3} z_2)+\frac{y}{2} (z_2 \partial_y z_2 + 
z_1 \partial_y z_1) +\rc
+ \frac{(2\pi\alpha')^2}{4} F_{MN}F^{MN} -\frac{64(2\pi\alpha')^2}{3^{\frac32}
} \epsilon^{abc}e_a^i
e_b^j e_c^k A_i \partial_{\alpha_j} A_k\Bigg) .\,
\label{fulllagr}
\eear
In this expression the indices $M,N=1,2\ldots 8$ label  the worldvolume coordinates on the D7-brane, while  $i,j,k=1,2,3$ label the angular coordinates $\alpha_i$ of the squashed sphere. 
$F$ is the abelian gauge field
strength on the probe D-brane, $F_{MN} = \partial_M A_N - \partial_N A_M$.
It is remarkable that fluctuations $z_i$ of the embedding  
and the gauge field $A_M$ do not mix at quadratic order in (\ref{fulllagr}).
We can now extract the spectrum of fluctuations from the equations of
motion that follow from (\ref{fulllagr}).  

There are two kinds of fluctuations on the probe brane: excitations of the
brane embedding in the target space and excitations of the worldvolume
gauge field. We will analyze each of these in turn below.

\subsection{Fluctuations of the embedding}
We first consider the two sets of oscillations transverse to 
the squashed three-sphere $\tilde S^3$ wrapped by the D7-brane,
including their Kaluza-Klein modes on $\tilde S^3$. These include
excitations of the $\theta$ 
coordinate around $\theta=0$. At $\theta=0$, we parametrized the
transverse fluctuations in terms of the non-singular coordinates 
$z_{1,2}$.  In the UV, the
$\theta$-mode (also called the slipping mode) is dual to a
dimesion-three, quark bilinear operator of the form $q^i {\tilde q}_i$
(cf.~eqn.~\eqn{compare}). 
At the Leigh-Strassler fixed point, the equations of motion for the $z_1$ and $z_2$ fields on the D7-brane are
\bear
\frac{\sqrt3 }{4\sqrt2 y^2} \partial^\mu \partial_\mu z_1+
\frac{y^2}{3} \partial_y^2 z_1 + \frac{5y}{3} \partial_y z_1 + \frac54 z_1 +
\nabla^2 z_1 + \frac12 \partial_{\alpha_3} z_2=0
\,\,,\rc
\frac{\sqrt3}{4\sqrt2 y^2} \partial^\mu \partial_\mu z_2+
\frac{y^2}{3} \partial_y^2 z_2 + \frac{5y}{3} \partial_y z_2 + \frac54 z_2 +
\nabla^2 z_2 - \frac12 \partial_{\alpha_3} z_1 =0
\,\,,
\label{eqszs}
\eear
where $\nabla^2$ stands for the laplacian on the $\tilde S^3$ as given in (\ref{laplacian}).
Inserting the scalar harmonics on the squashed sphere (see Appendix
\ref{app:harmonics}), we can easily decouple the equations
for the two modes by writing
\be
z_1 = e^{iq\cdot x} {\cal Z}_\pm (y) Y^l_{mn} (\alpha)
\,\,,\qquad
z_2 = \pm i\, z_1\,. 
\ee
Here $q^\mu$ is the four-momentum in the spacetime directions along the boundary gauge theory. We then find
\be
 -q^2 \frac{\sqrt3 }{4\sqrt2 y^2}  {\cal Z}_\pm (y)+
\frac{y^2}{3} \partial_y^2 {\cal Z}_\pm (y) + \frac{5y}{3} \partial_y {\cal Z}_\pm (y) + 
\frac54 {\cal Z}_\pm (y) -(\lambda \mp \frac{n}{2}) {\cal Z}_\pm (y) =0
\,\,.
\label{eqscalar}
\ee
$\lambda$ is the eigenvalue written in (\ref{lambdascalar}) with $a^2=1$,  $b^2= 4/3$, so that
\be
\lambda=l(l+1) - \frac14 n^2\,.
\label{lambdascalar2}
\ee
 In order to find the dimensions of the operators dual to these
 oscillation modes, we follow the
standard AdS/CFT rules. First we determine the form of the solution of (\ref{eqscalar}) near the boundary $y\to \infty$, with the result
\be
{\cal Z}_\pm (y) \sim c_1 \, y^{-2 + \frac12 \sqrt{1+12 (\lambda \mp n/2)}} + 
c_2 \, y^{-2 - \frac12 \sqrt{1+12 (\lambda \mp n/2)}} \,.
\ee
From this asymptotic expansion we read off the conformal dimensions of the dual field theory operators:
\be
\Delta^z_\pm = 2 + \frac{1}{2} \sqrt{1 + 12 l(l+1) - 3 n^2 \mp 6n}\sac l=0, \frac{1}{2}, 1, \ldots \sac |n|\leq l \,.
\ee
We see that the $l=0$ mode dual to the quark bilinear $q^i {\tilde q}_i$
has dimension $5/2$ at the IR fixed point. The $SU(2)$
 isometry of $\tilde S^3$ is dual to the global $SU(2)$ symmetry of
 the Leigh-Strassler SCFT, under which the chiral multiplets $\Phi_1$
 and $\Phi_2$ transform as a doublet. 
Therefore the $SU(2)$ quantum
number $l$ specifies appropriate insertions of the components of  
$\Phi_{1,2}$ into the fermion
 bilinear yielding operators in an irreducible 
representation of $SU(2)$. Interestingly, 
for $n=\pm l$ and for $n=\pm l \mp 1$, the dimensions become rational numbers:
\be
\Delta^z_\pm\big|_{n=\pm l} = \frac52 + \frac32 l\,\,,\qquad
\Delta^z_\pm\big|_{n=\pm l\mp 1} = 3 + \frac32 l\,\,.
\ee

\subsection{Fluctuations of the worldvolume gauge field}
We now turn to operators dual to fluctuations of the worldvolume gauge field on the probe D7-brane. To study these,  we will first partially fix the gauge by demanding
\be
\eta^{\mu \nu} \partial_\mu A_\nu = 0\,.
\label{transverse}
\ee
Inserting this gauge choice into the equations of motion, we obtain
relatively simpler expressions. 
There are three distinct types of gauge field components: the radial
mode $A_y$, components $A_i$ along the internal directions of the
$\tilde S^3$ wrapped by the D7-brane, 
and the four-vector $A_\mu$. The only contribution to the $A_y$
equation of motion comes from the $F_{MN}F^{MN}$ term, and it takes
the form  
\be
\frac{\sqrt3 }{4\sqrt2 y^2} \partial^\mu \partial_\mu A_y
+\nabla^2 A_y - \partial_y\left[ \frac{1}{\sqrt{\tilde g_3}}\partial_i \left(
\sqrt{\tilde g_3} \, \tilde g^{ij}  A_j \right)\right]=0\,.
\label{eqforAy}
\ee
Similary, the equation of motion for $A_\nu$ is
\be
\frac{\sqrt3\, }{4 \sqrt2 \,y^2}\partial^\mu \partial_\mu A_\nu + \frac{1}{3y}
\partial_y \left( y^3 \partial_y A_\nu \right) + \nabla^2  A_\nu 
-\partial_\nu \left[ \frac{1}{3y} \partial_y (y^3 A_y)+
\frac{1}{\sqrt{\tilde g_3}}\partial_i \left(
\sqrt{\tilde g_3} \, \tilde g^{ij}  A_j \right)
\right]
=0\,.
\label{eqforAnu1}
\ee
Finally, the linearized equation for $A_i$ reads
\bear
&&\tilde g^{mj}\left[ \frac{\sqrt3 }{4\sqrt2 y^2}\partial^\mu \partial_\mu A_j
+\frac{1}{3y^3} \partial_y(y^5 \partial_y A_j)- \frac{1}{3y^3}
\partial_y (y^5 \partial_j A_y)\right]
-\frac{4}{\sqrt3 \sqrt{\tilde g_3}}\epsilon^{mjk} \partial_j A_k +\rc
&&+\frac{1}{\sqrt{\tilde g_3}}\partial_j \left[ \sqrt{\tilde g_3}
\tilde g^{jk} \tilde g^{ml} (\partial_k A_l - \partial_l A_k) \right]=0\,.
\label{eqforAi1}
\eear
This can be written in a simpler form after contracting with 
$\tilde g_{im}$:
\be
\frac{\sqrt3 }{4\sqrt2 y^2}\partial^\mu \partial_\mu A_i
+\frac{1}{3y^3} \partial_y(y^5 \partial_y A_i)- \frac{1}{3y^3}
\partial_y (y^5 \partial_i A_y) - {\cal O}_1({\cal O}_1 A_i) - \frac{4}{\sqrt3}
{\cal O}_1 A_i=0\,\,,
\label{eqforAi}
\ee
where the differential operator ${\cal O}_1$  is defined in (\ref{O1O2})

Below, we analyze the spectrum of these three kinds of gauge modes,
and classify them using the notation of \cite{Kruczenski:2003be}. 

\subsubsection*{Type I modes}
We set $A_\mu=A_y=0$ and write $A_i$ in terms of the vector spherical
harmonics on the $\tilde S^3$ described in
(\ref{vectorharm1})-(\ref{vpm}) as 
\be
A_i = h_1^\pm (y) e^{iq\cdot x} \left({\bf Y}^\pm\right)_i \,.
\label{ansatzforAi}
\ee
With the ansatz (\ref{ansatzforAi}), the only non-trivial equation of
motion is (\ref{eqforAi}), which becomes 
\bear
-q^2\frac{\sqrt3}{4\sqrt2 y^2} h_1^\pm (y) 
+\frac{1}{3y^3} \partial_y(y^5 \partial_y h_1^\pm (y) )
-v_\pm^2 h_1^\pm (y)  -\frac{4}{\sqrt3}v_\pm \, h_1^\pm (y) 
=0\,.
\label{Aimodes}
\eear
%DDD New sentence below MMM
As usual, the contribution from the four-momentum (the $q^2$ term) is
subleading near the boundary ($y \to \infty$) and therefore this term
can be neglected in order to compute the near-boundary solution, which
takes the form 
\be
h_1^\pm (y) \sim c_1 y^{-4-\sqrt3 v_\pm} + c_2 y^{\sqrt3 v_\pm} \,.
\ee
The resulting conformal dimensions of the dual operators are then
\be
\Delta^I_+ = 4 + \sqrt3 v_+ \,\,,\qquad {\rm and} \,\, \qquad \Delta^I_- = - \sqrt3 v_-\,\,.
\ee
Substituting in the eigenvalues $v_{\pm}$ from (\ref{vpm}) we finally find
\bear
\Delta^I_+ &=& 5 + \frac12 \sqrt{4+12 l +12l^2 -3n^2} \sac
l=0,\frac12, 1, \ldots \sac |n|\leq l+1 \,,\rc
\Delta^I_- &=& -1+ \frac12 \sqrt{4+12l+12l^2 - 3n^2} \sac 
l=1,\frac32, 2, \ldots \sac |n|\leq l-1 \,.
\eear
Notice that $\Delta^I_+$ becomes rational for some particular values of $n$
\be
\Delta^I_+|_{|n|=l+1}=\frac{11}{2}+\frac32 l\,\,,\qquad
\Delta^I_+|_{|n|=l}= 6+\frac32 l\,\,.
\ee

\subsubsection*{Type II modes}
Next we turn to the Type II modes which are dual to vector operators in the gauge theory, with $A_y=A_i=0$ and
\be
 A_\mu=\xi_\mu 
e^{iq\cdot x} h_2(y) Y^l_{mn} \,,
\ee
for a constant transverse polarization vector $\xi_\mu$. These modes are vectors in the dual
field theory, as opposed to the rest of the bosonic fluctuations, which are all scalar operators.
It is then clear that the
only non-trivial equation is (\ref{eqforAnu1}), which reduces to
\be
-q^2 \frac{\sqrt3}{4\sqrt2 y^2}h_2(y)+
\frac{1}{3y}
\partial_y \left( y^3 \partial_y h_2(y) \right) -\lambda \,  h_2(y)=0\,.
\ee
For large $y$, the asymptotic solutions are 
\be
h_2 \sim c_1 y^{-1-\sqrt{1+3\lambda}} + c_2 y^{-1+\sqrt{1+3\lambda}}
\ee
and thus, using (\ref{lambdascalar2}), we arrive at
\be
 \Delta^{II}=2 + \sqrt{1+3l(l+1)-3\frac{n^2}{4}} \sac
 l=0,\frac12, 1, \ldots \sac |n|\leq l \,.
 \label{Delta2}
\ee
For a fixed $l$, the dimension is minimal when $n=\pm l$ and in fact
$\Delta^{II}$ becomes a rational number in these cases:
\be
\Delta^{II}|_{|n|=l}  = 3 + \frac32 l\,\,.
\ee

\subsubsection*{Type III modes}

Finally we look at the gauge field fluctuations with $A_\mu=0$  and
\be
A_y= e^{iq\cdot x} h_3(y) Y^l_{mn} \sac
A_i = e^{iq\cdot x} \tilde h_3(y) \partial_i Y^l_{mn}\,\,\,.
\ee
Equation (\ref{eqforAnu1}) fixes $\tilde h_3$ in terms of $h_3$ as
\be
\tilde h_3= \frac{1}{3y\lambda}\partial_y(y^3 h_3)\,\,,
\label{tildeh3sol}
\ee
and the equation of motion for $h_3(y)$ follows from (\ref{eqforAy}):
\be
-\frac{\sqrt3 }{4\sqrt2 y^2}q^2  h_3+
\partial_y \left(\frac{1}{3y}\partial_y(y^3 h_3) \right)-\lambda \,h_3 =0\,\,,
\ee
while Eqn.~(\ref{eqforAi}) is automatically satisfied.
At large $y$, the solution is 
\be
h_3 \sim c_1 y^{-2-\sqrt{1+3\lambda}} + c_2 y^{-2+\sqrt{1+3\lambda}}
\ee
and thus
\be
 \Delta^{III}=  2 + \sqrt{1+3l(l+1)-3\frac{n^2}{4}} \sac
 l=\frac12,1, \frac32, \dots \sac |n|\leq l \,.
 \label{Delta3}
 \ee
We point out that  the trivial mode on the sphere,
$\lambda=0$, is not allowed due to the factor of $\lambda$ in the denominator of Eqn.~(\ref{tildeh3sol}).
This reflects the fact that $l=0$ is excluded in 
Eqn.~\eqn{trivialvectorharm}. Apart from the absence of the $l=0$ mode, the eigenvalues (\ref{Delta3})
coincide with those for the vectors in Eqn.~(\ref{Delta2}).

\subsection{A discussion on the dimensions}

We now compare the spectrum of  operator dimensions at the IR fixed
point to the UV dimensions so that we can identify the anomalous
dimensions gained along the RG flow. This is easily done by comparing
with the results of \cite{Aharony:1998xz, Kruczenski:2003be}, where
the dimensions of  
these operators at the UV fixed point (${\cal N}=4$ SYM coupled to massless ${\cal N}=2$ matter) were computed. 
In order to translate the results of \cite{Kruczenski:2003be} to our language one must make the the replacement
$l\to 2l $ for all modes, as well as a subsequent shift of $l$ by $\pm
1$ for the type I $\pm$ modes. Note also that the discrete label $n$
used in the analysis of \cite{Kruczenski:2003be} is unrelated to the
quantum number $n$ we are using here. 
The final result is 
\begin{displaymath}
\Delta^{z,UV}_{\pm} =\Delta^{II,UV} =\Delta^{III,UV} = 3+2l
\sac \Delta^{I,UV}_+ = 6 + 2l
\sac \Delta^{I,UV}_- =  2l \,.
\end{displaymath}
The main points to note are as follows. First, the conformal
dimensions at the IR fixed point are irrational numbers in general. At
the UV fixed point all dimensions are integer-valued because the
D7-brane modes fall into short multiplets of the four-dimensional
${\cal N}=2$ super\emph{conformal} algebra, and hence their scaling
dimensions are completely determined by their transformation
properties under the $SU(2)\times U(1)$ R-symmetry of the UV
theory. In contrast, the Pilch-Warner flow and the IR SCFT are only
invariant under ${\cal N}=1$ supersymmetry. For this reason, the only
constraint on scaling dimensions at the IR fixed point arises from
holomorphy and from the charges carried by (anti)-holomorphic/chiral
operators under the IR $U(1)_R$ symmetry. Consequently, most of the
operators corresponding to fluctuations of the D7-brane have
non-trivial, irrational anomalous dimensions. In particular, all the
operators which are non-holomorphic or non-chiral in the  
${\cal N}=1$ sense will likely acquire irrational anomalous dimensions.

Second, simple inspection shows that all the IR operators have a
smaller dimension than their UV counterparts, except for the lowest
type II and type I+ modes, whose dimensions 
\be
\Delta^{II}|_{l=0} = 3 \,\,\,\,\,\,\,\,\, \mbox{and} \,\,\,\,\,\,\,\,\, 
\Delta^{I}_+|_{l=0,n=0} = 6 
\ee
remain unchanged. The former is protected because the $l=0$ harmonic of the worldvolume gauge field $A_\mu$ is dual to a conserved $U(1)_B$ current in the gauge theory,
and conserved vector currents must have scaling dimension 3. 
Under $U(1)_B$, which commutes with the supersymmetry generators, the superfields $Q^i$ and $\tilde
Q_i$ transform with charges $+1$ and $-1$ respectively. 
The reason why the dimension of the operator dual to the type I+ mode is not renormalized is unclear to us.
All operators other than the two above appear to acquire order-unity, negative anomalous dimensions along the flow.

Third, the dimension of the lowest-lying scalars describing fluctuations of the embedding are reduced from 3 in the UV to 
$\Delta^{z}_{\pm}|_{l=0} = 5/2$ in the IR. Among these is the `slipping mode' which, as discussed previously, is dual to the quark bilinear operator $q^i {\tilde q}_i$
that typically develops a non-trivial expectation value when supersymmetry is broken \cite{Babington:2003vm,Kruczenski:2003uq}. 

Finally, it is not entirely clear to us why some of the dimensions
become rational for special choices of $n$ with fixed $l$. However, it
appears that in all such cases the 
simplified dependence on $l$ can be accounted for by insertions of the
adjoint fields $\Phi_{1,2}$ in the mesonic operators. Indeed, these
fields effectively have a scaling dimension of $3/4$ at the fixed
point and  
carry `spin 1/2' under the $SU(2)$ symmetry of the theory. Thus we see
that $n$ insertions of the form  
$Q \Phi^n{\tilde Q}$ amount to increasing $l$ by 
$n/2$.

%%%%%%%%%%
%\newpage
\section{Spectrum and a light dilaton}
\label{Sec:3}
In this section we take a first step towards analyzing the
Pilch-Warner flow away from the IR fixed point and studying possible
deformations driving the theory away from the walking
regime. Motivated by hard-wall approaches to confining dynamics, we
cut off the Pilch-Warner geometry in the infrared by hand. 
Because of technical convenience, we also introduce a UV cut-off, but this could be easily removed. 
As a consequence of the IR cut-off the spectrum becomes discrete and
gapped. Our primary goal is to carefully compute this spectrum and to
specifically look for a light dilaton-like state.  
We focus on the supergravity spectrum. In Section \ref{sec:discuss} we offer a preliminary discussion of possible studies of D7-brane dynamics in the cut-off geometry, but we leave a detailed investigation for the future.

We consider fluctuations of  the system of scalars coupled to gravity
 around the five-dimensional background dual to the LS flow. In the
 language of Section \ref{Sec:2} these are flows that start near $F_U$
 and  evolve towards the fixed point $F_I$. In practice, due to finite
 numerical precision, the 
 flows never actually get to $F_I$; they can be made to get
 arbitrarily close to it, but eventually they all deviate away and 
 end at a singularity.  The singularity is actually of a `bad' type
 (see section (\ref{sec:fp}) and Appendix \ref{app:sing}) and its
 supergravity description is problematic. However, this will not be an
 obstacle because we will cut off the geometry in the IR well above
 the scale where any pathological behaviour sets in. 

In order to compute the spectrum we will follow the algorithm
and the notation of Ref.~\cite{EP}. In particular, we will compute the
spectrum by restricting  the radial direction to be compact, $r_1 < r
< r_2$, with hard IR and UV cut-offs $r_1$ and $r_2$,
respectively. This in turn means that we will need to add  
localized boundary actions at the ends of the space, $r=r_i$, and
choose appropriate boundary conditions  for the fluctuations. We will
always choose $r_1\gg r_0$, where $r_0$ is the end-of-space where all
flows passing arbitrarily close to $F_I$ eventually end up. We also
choose $r_1>r_I$, where $r_I$ is the scale below 
 which the flow drifts away from the IR fixed point towards $r_0$. In
 a completely consistent walking scenario $r_I$ would represent the
 energy scale $\Lambda_I$ depicted in Fig. \ref{Fig:coupling}. As the
 radial dimension is compact,  
the resulting spectrum will be discrete.
In this way, we will study the effects of the walking region 
on the spectrum.  Due to our choice of IR cut-off the results are
completely insensitive to any singularity sitting in the deep IR. 

The meaning of the two cut-offs needs to be clarified. Since  the
geometry is asymptotically AdS, the UV cut-off could simply be removed
by taking the limit $r_2\rightarrow +\infty$  (after inclusion of
appropriate counter-terms, along the lines of holographic
renormalization~\cite{HR}). 
We will not do so because our results are not qualitatively affected by this procedure. 
The role of the IR cut-off is more interesting. This could be thought
of as an IR regulator on the dual field theory. The fact that we
choose $r_1>r_I$ means that  
the theory is effectively conformal in the IR all the way down to
$r_1$, at which point confinement abruptly takes place, as modeled by
the hard wall.  
Of course, the theory is not conformal at all scales above
$r_1$. There is a dynamically-generated scale $r_{\ast}$ between the
two cut-offs where the transition between the UV and the IR
approximately-conformal dynamics takes place. The difference $r_\ast -
r_1$ can be roughly thought of as the size of the walking region.  

We are interested in the dependence of physical observables on
$r_\ast$. Indeed, although the discretization of the spectrum would
survive in a complete string model in which confinement arises truly
dynamically,\footnote{Note that this would require additional
  deformations of the field theory by relevant operators which deform
  the PW solution in the deep-IR region.} the dependence of the
spectrum on the scale $r_1$ is likely to change. In contrast, the
distortion of the spectrum due to the $r_{\ast}$ scale, provided this
is  far above the confining scale, would presumably remain. In
summary, we wish to understand what changes in the 
spectrum when we keep the IR and UV cut-offs fixed 
but vary the boundary conditions so that $r_{\ast}$ changes.
In particular, we will see that this has a crucial effect on the 
lightest state of the spectrum, which we would like to identify with a pseudo-dilaton.

\subsection{Gauge-invariant fluctuations and numerical results}

In order to determines the spectrum of scalar particles in the gauge
theory we must consider fluctuations of the five-dimensional
supergravity scalars with well-defined four-momentum, $\phi^a (q,r)$,
subject to appropriate boundary conditions at $r_i$. The particle
masses are then the values of  $M^2 = q^2$ for which solutions
exist. This procedure is complicated by the fact that the fluctuations
of the scalars source those of the five-dimensional metric. In other
words, all the fluctuations are coupled.  Here we will deal with this
difficulty following the formalism of~\cite{BHM,EP}, to which we refer
the reader for further details. The idea is to write the equations of
motion for the fluctuations in terms of appropiate gauge-invariant
combinations of the scalar and metric fluctuations. In terms of these
new scalar fields, which we denote as $\mathfrak{a}^a (q,r)$ following
the references above, one is left (in the scalar sector) with a
system of two coupled, second-order, linear differential equations. 

In our present case, the formalism is greatly simplified for the
following reasons:
\begin{itemize}
\item The five-dimensional sigma-model metric is particularly simple: the associated  connection is trivial 
with ${\cal G}^a_{\,\,bc}=0$.
\item The supergravity model is endowed with a superpotential $W$ (\ref{eq:superpot}), 
and hence the equations can be written in
  terms of $W$ and its field derivatives. 
\item There are only two active scalars in the system.
\end{itemize}

We also make another simplification. In defining the boundary actions,
there is some freedom 
as to the choice of a set of couplings, which can be thought of as
effective localized mass matrices 
for the sigma-model scalars. In general, the spectrum depends on these matrices.
Furthermore, there always exist choices of such mass terms that makes
one or more of the scalar  excitations exactly massless. As we are
mainly interested in understanding whether a light scalar excitation
is admitted by the backgrounds we are considering, we take the most  
{\it conservative} possible attitude, and take all of these mass terms
to diverge. 
This choice is conservative in the sense that if we find (as we will)
a light state, 
this would still be a light state for any other choice of the boundary
masses, and hence it can be taken as a physical  result as opposed to
an artifact associated to specific choice of boundary conditions. 

The equations we have to solve involve two
scalar fields $\mathfrak{a}^a$ related to the fluctuations of the
sigma-model scalars through  
\beqs
\mathfrak{a}^a&\equiv&\varphi^a+\frac{W^a}{4W} h\,,
\eeqs
where $h$ is the trace of the four-dimensional
metric fluctuations. The field derivative $W^a$ is evaluated on the
classical background, and is defined by  
\be
W^a=G^{ab} \frac{\partial W}{\partial \phi^a} 
=\partial_r \phi^a \,.
\ee
The equations can then be written as
\SP{\label{Eq:diffeqWN}
        \Bigg[ e^{-4A} \left( \delta^a_b \partial_r + N^a_{\,\,\,\,b} \right)
        e^{4A}\left( \delta^b_c \partial_r - N^b_{\,\,\,\,c} \right) +
        \delta^a_c e^{-2A} \Box \Bigg] \mathfrak{a}^c = 0\,, 
}
and the boundary conditions take the form 
\SP{\label{Eq:BCWinfty}
        &\left[ e^{2A} \Box^{-1}\frac{W^c W_d}{W}\right]
        \left(\delta^d_{\,\,\,\,b} \partial_r - N^d_{\,\,\,\,b}
        \right)\mathfrak a^b \Big|_{r_i}  
        =   \delta^c_{\,\,\,\,b} \mathfrak a^b \Big|_{r_i}\,,
}
where $\Box=-K^2=q^2$ is the four-dimensional momentum
and 
\beqs 
N^d_{\,\,b}&\equiv&
G^{dc}\partial_c\partial_b W - \frac{W^d W_b}{W} \,.
\eeqs
All the functions $N$, $W$, $W_c$, $A$ and $W^c_{\,\,\,\,b}$ are
evaluated on the classical background (which is known numerically).  

As anticipated, the system reduces to two coupled, second-order,
linear equations in the functions $\mathfrak{a}^a(q,r)$,  
subject to two sets of boundary conditions in the IR and UV. Note that
the boundary conditions 
  have the form of generalized Neumann
boundary conditions, involving both the field and its derivative. However, note 
that in the limit in which the boundaries approach a fixed point ($W_a=0$)
with non-trivial AdS curvature ($W<0$), the left-hand side of these expressions vanishes, and the boundary conditions  
reduce effectively to Dirichlet (provided $q^2$ is not small).

\begin{figure}[h]
\begin{center}
\begin{picture}(430,360)
\put(0,170){\includegraphics[width=7.0cm]{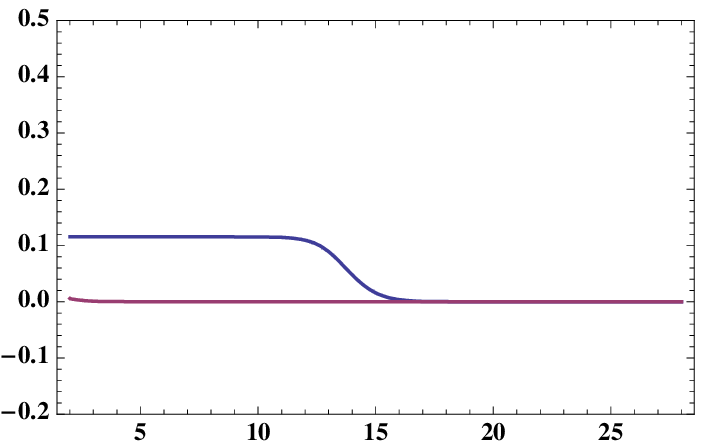}}
\put(220,170){\includegraphics[width=7cm]{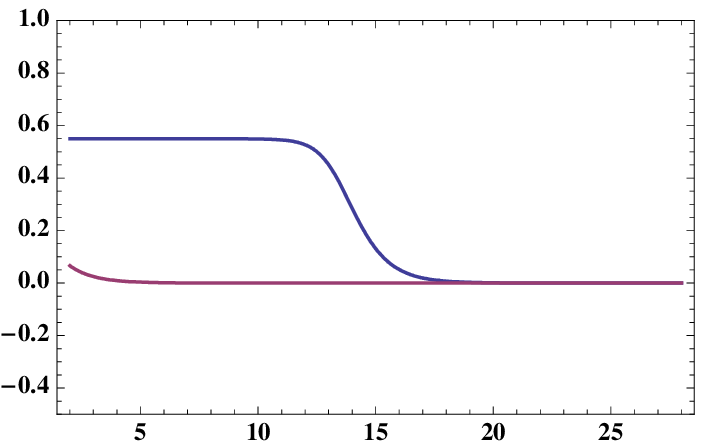}}
\put(0,0){\includegraphics[width=7cm]{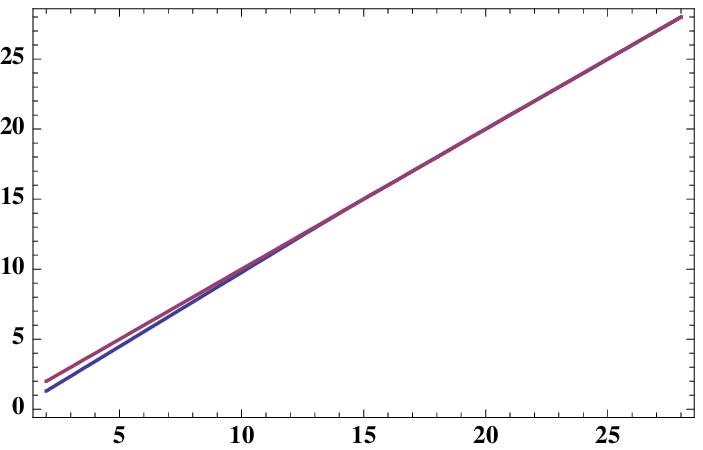}}
\put(220,0){\includegraphics[width=7cm]{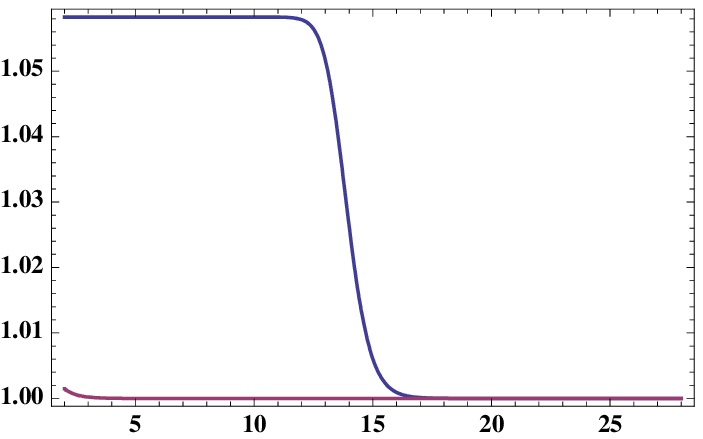}}
\put(-8,280){${\alpha}$}
\put(-8,110){${A}$}
\put(213,280){${\chi}$}
\put(213,110){${A}^{\prime}$}
\put(400,165){$r$}
\put(400,-5){$r$}
\put(170,165){$r$}
\put(170,-5){$r$}
\end{picture} 
\caption{The functions  ${\alpha}$, ${\chi}$, ${A}$ and ${A}^{\prime}$ defining the backgrounds used in the 
calculation of the spectrum. In blue the case $r_{\ast}\simeq 14$ and in red the case $r_{\ast}\simeq 0.3$, 
plotted in the range $r_1=2<r<r_2=28$
used in computing the spectrum. 
 }
\label{Fig:bg}
\end{center}
\end{figure}

\begin{figure}[htpb]
\begin{center}
\begin{picture}(430,260)
%\put(0,0){\includegraphics[width=12.8cm]{spectrumplotmrho2806_gr1.eps}}
\put(0,0){\includegraphics[width=14cm]{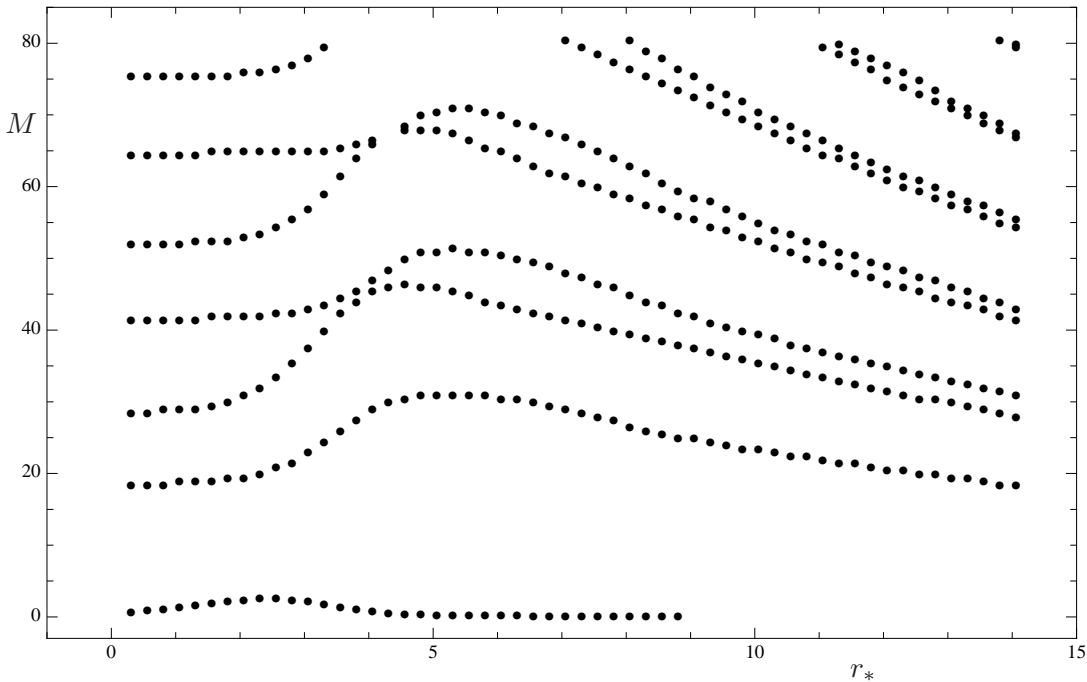}}
\put(-6,200){$M$}
\put(310,-5){$r_{\ast}$}
\end{picture} 
\caption{Numerical results for the spectrum, $r_1=2$, $r_2=28$,
as a function of $r_{\ast}$ (see text).
 }
\label{Fig:Spectrum}
\end{center}
\end{figure}

\begin{figure}[htpb]
\begin{center}
\begin{picture}(430,260)
%\put(0,0){\includegraphics[width=12.8cm]{spectrumplotmrho2806log_gr1.eps}}
%\put(0,0){\includegraphics[width=12.8cm]{plotscalarslog_gr1.eps}}
\put(0,0){\includegraphics[width=14cm]{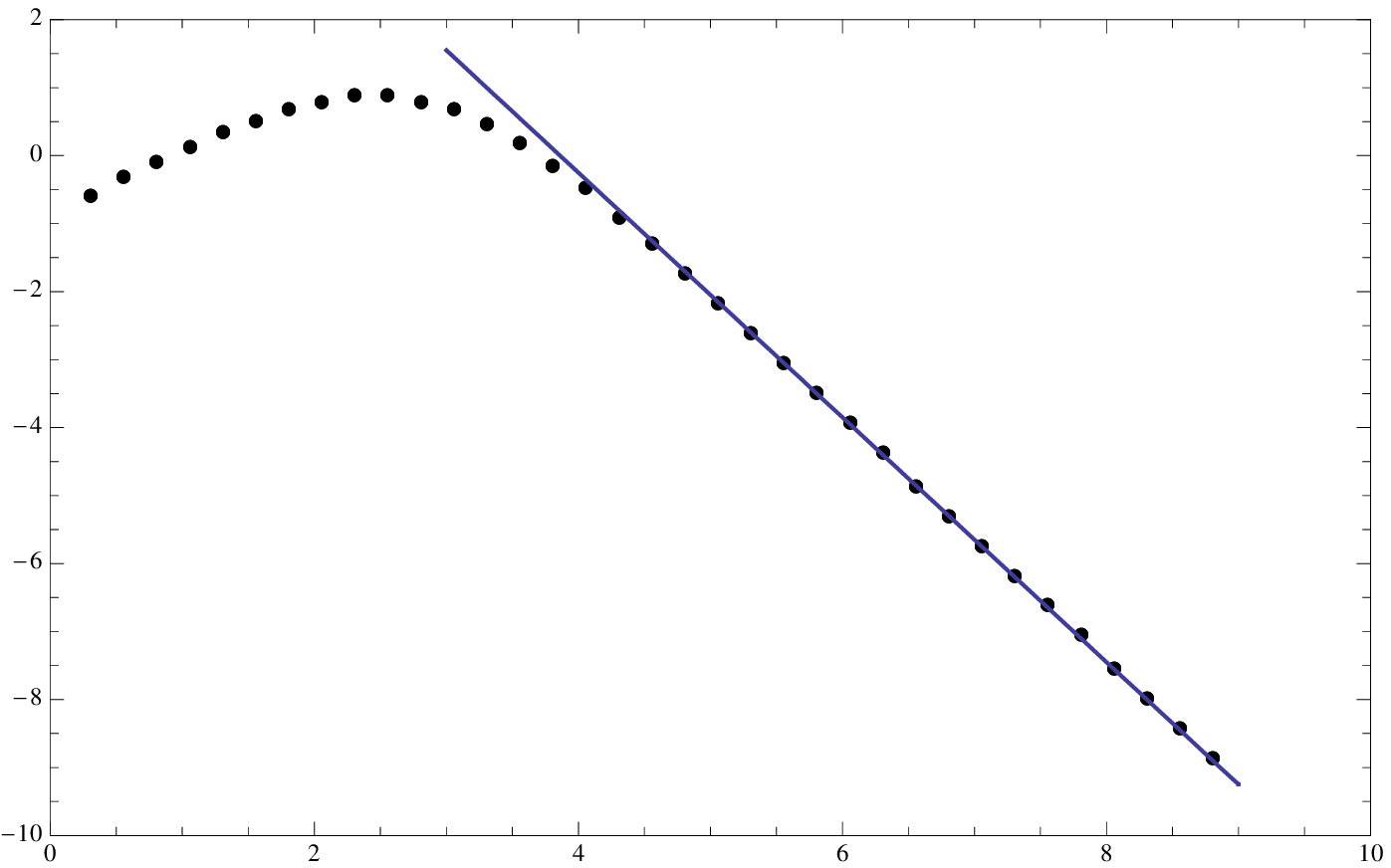}}
\put(-11,220){$\ln M$}
\put(330,-5){$r_{\ast}$}
\end{picture} 
\caption{Numerical results for the spectrum, for $r_1=2$, $r_2=28$,
as a function of $r_{\ast}$ (see text). Only the lightest state is shown, together with 
the semi-analytical expectation.
 }
\label{Fig:Spectrumlog}
\end{center}
\end{figure}

In order to determine the spectrum we first numerically generate a set
backgrounds for the five-dimensional metric and scalars that describe
LS flows from $F_U$ to $F_I$ in the language of Section
\ref{Sec:2}. For all these flows we choose the cut-offs to be $r_1=2$
and $r_2=28$, making sure that the singularity always appears at some  
$r_0\ll r_1$, but vary the boundary conditions so that 
$r_{\ast}$ changes from flow to flow. We define $r_{\ast}$ as the
value of $r$ for which the  scalar ${\chi}=( \ln 3 )/4$, half-way
between its asymptotic UV and IR values --- see Eqs.~\eqn{Eq:UV} and
\eqn{Eq:IR}. We could have equivalently fixed $r_\ast$ and changed the
position of the cut-offs, but this is less convenient
numerically. Either way, the size of the walking region changes from
flow to flow. 
We also choose the integration constant in the warp factor so that the
UV-asymptotic behavior of $A(r)$ is identical for all the flows. The
functions that determine the background solution, ${\alpha}(r)$,
${\chi}(r)$, and ${A}(r)$, are shown in Fig.~\ref{Fig:bg} for two
different flows with $r_{\ast}\simeq 0.3$ and $r_{\ast}\simeq
14$. With the flows in hand, we study scalar fluctuations as described
above and detrmine the spectrum of scalar particles for each flow
using the mid-point determinant method. The results are shown in
Fig.~\ref{Fig:Spectrum} and Fig.~\ref{Fig:Spectrumlog}. 

\subsection{Interpretation of the results}

In order to understand some of the features that emerge from the
numerical results it is useful to consider a simplified model in which
the solution is a small deviation from AdS. The details are discussed
in Appendix \ref{simple}. Here we will just describe the salient
aspects of the numerical results and refer the reader to the appendix
at some specific points:  
\begin{itemize}
\item For small values of $r_{\ast}$, the spectrum agrees with the
  limiting case in which the background is purely AdS, with unit
  curvature $L=1$. 
Specifically, precisely as in the RS1 case, there is a massless
dilaton accompanied 
by a tower of equally spaced KK-modes.
The reason for this is simply that we are effectively turning off the
deformations of the CFT living at the UV fixed point. 
This can be easily understood in terms  of $\alpha(r_1)$ and
$\chi(r_1)$, which, instead of approaching their IR-asymptotic  
values, are simply suppressed. In addition, the background has a constant $A^{\prime}=1$.
This limit is not very interesting. 

\item For large values of $r_{\ast}$, the heavy KK states form pairs,  for accidental reasons that are explained
in Appendix \ref{simple}. When comparing the  analytical example with the numerics one should replace the UV cut-off  $r_2$ in the example, with $r_{\ast}$.

\item A cross-over behavior appears at intermediate values of $r_{\ast}$. This is not surprising because 
the general spectrum has to interpolate between the two spectra obtained at the two fixed points.

\item The two lightest states do not show the pairing displayed by the
  heavy states. One of them is in fact very light for any $r_{\ast}$ and exhibits a strong dependence
on $r_{\ast}$ that is most interesting for our purposes. We will
  comment on this in more detail below. The other state is just the
  lightest of the $M_+$ states (see Appendix  \ref{simple}) computed
  in the vicinity of the IR fixed point. 

\item The entire spectrum is mildly suppressed at relatively larger
  values of $r_{\ast}$. This is not an interesting  
physical feature; it is an artifact of the way in which we set up the calculation.
In particular, we chose all the backgrounds to asymptote to AdS with
$L=1$ in the far-UV. 
Since in the IR (below $r_{\ast}$) the background is again AdS, but
with a different curvature, fixing the IR cut-off at $r_1=2$
corresponds to different choices of the physical IR scale.
Hence the spectra computed at different values of $r_{\ast}$ are
  normalized with slightly different  IR scales. 
For the KK modes, which have mass controlled by $e^{r_1}$, this
results in a rescaling by a factor of $e^{(1-1/L_{IR})r_{\ast}}$.
For the light dilaton, the rescaling is more complicated, and we will explain it later.
\end{itemize}

Before we focus on the mass of the lightest state, we perform some semi-quantitative checks.
The first heavy KK-mode should have a mass (from the approximate results in Appendix  \ref{simple})
\beqs
M_{11}&\simeq&\frac{2.4}{L_1}\,\equiv\,2.4\,\,e^{r_I}\,\simeq\,18
\eeqs
in the case in which the background is given by the UV fixed-point.
This is in agreement with what we found for small values of $r_{\ast}$.
The spacing between  KK-modes should be
\beqs
\Delta M &\simeq & \frac{\pi}{L_1}\,\simeq \, 23\,,
\eeqs
and hence $M_{1i}\simeq 18, 41, 64, 87\,\cdots$, which is again in
remarkable agreement with the numerical results. 
The second such tower starts at 
\beqs
M_{21}&\simeq&\frac{5\pi}{4 L_1}\,\simeq\,29\,,
\eeqs
and yields $M_{2i}=29,52,75\,\cdots$, also in agreement with the numerical results.
The two towers correspond to the fluctuations of the two scalar
fields  which 
have five-dimensional masses $L^2 M_\mt{scalar}^2=-3,-4$ respectively --- see Eqn.~\eqn{scalar}.

One can  perform analogous tests at large values of $r_{\ast}$, and again very good agreement is found
with the results obtained by expanding around the IR fixed-point.

Let us now turn to  the lightest state in the spectrum, which  becomes exactly massless in the $r_{\ast}\rightarrow \infty$ limit. Firstly, we point out that the behavior at low values of $r_{\ast}$ is
uninteresting: this parametric regime corresponds to placing the IR
cut-off at a very high scale,
freezing out the deformations of the CFT living at the UV fixed point.
More interesting is what happens when $r_{\ast}\gsim 3$. The fact that the mass appears to be suppressed by the $r_{\ast}$ 
scale signals the following interpretation. While it is undoubtedly true that the dual
theory is not conformal on all scales, the
emergence of the light state is due to the fact that an approximate
scale invariance is present in the walking region $r<r_{\ast}$. 
This scale invariance is approximate as it is 
explicitly broken by an irrelevant coupling. The spontaneous breaking
is the result of our crude attempt to replace the IR with a hard-wall cut-off.
This  must indeed be the case, since in this region the RG flow is 
approaching the IR fixed point, 
and it would not do so if relevant deformations were present.

We can make this statement more quantitative, explicitly using our results.
Notice that, because the background was obtained numerically, it does not reach
the IR fixed point exactly: both the VEV and the irrelevant coupling
are present. 
However, the former is effectively suppressed by the fact that we are
close to the fixed point 
(and we always choose $r_1$ to be far from any deep-IR singularity).
In the present case, the model is near-AdS only up to $r_{\ast}$,
and hence in the expression Eq. (\ref{eq:massformula}), we should
make the replacement $r_2\rightarrow r_{\ast}$. 
In order to explain the numerical results,
we also  need a correction factor for every appearance of $r_1$ in the
final mass formula. In particular, we expect  the lightest mass $M$ to scale as
 \beqs
 M&\propto& \exp\left[
 \left(1 - \sqrt{7}+ \sqrt{7} (1 - 1/L_{IR})\right) r_{\ast}
 \right]\,,
 \eeqs
which is in very good agreement with the numerical results, as shown
in Fig.~\ref{Fig:Spectrumlog}. 
Notice that in the figure an overall constant has been chosen to match
the data because  we are only interested in the functional dependence
on $r_{\ast}$, the only physically 
meaningful scale in this calculation.

\subsection{Comparison with Goldberger-Wise}

We close with a slight digression aimed at
readers who are familiar with the 
Goldberger-Wise mechanism~\cite{GW} (GW) and recognize that the formal
treatment we have followed in this section is similar to what is done
in that context. The main point we want to explain is that, while
there are certainly analogies with the GW mechanism at the technical
level, there are also important physical differences. 

The basic physical problem that the GW mechanism is designed to solve is a  fine-tuning problem
in the basic set-up of five-dimensional effective models in which an AdS bulk is bounded by two 
physical cut-offs (boundaries), such as~\cite{RS1}. The presence
of such boundaries, and the fact that they support physical degrees of
freedom, 
means that the theory is not scale invariant, and hence leaves open
the question of why we should be allowed to  
assume that the two boundaries are parametrically separated. At face
value this is a perfectly valid assumption as 
long as no other degrees of freedom are present, but it is a
fine-tuned choice if we associate the two boundaries to the  
scales characterizing electro-weak and gravity interactions.
The GW mechanism is implemented by adding a five-dimensional scalar
with a small five-dimensional mass. 
Effectively, this can be interpreted in terms of a quasi-marginal
deformation being added to the CFT 
dual to the AdS bulk theory. The marginal character of the deformation
means that the separation 
of scales can be explained dynamically in terms of the exponential of
the ratio between VEVs evaluated at the two boundaries, which can be
chosen to be of the same 
order of magnitude without fine-tuning any of the 
parameters in the initial action.
In this sense, the RS1 model, supplemented by the GW mechanism, is a
brilliant effective field theory solution to the 
electro-weak hierarchy problem.
Interestingly, this scenario leads to the presence of a light
pseudo-dilaton in the spectrum~\cite{GWspectrum}. 

What we have done in this section is substantially different for three important reasons.
Firstly, the bulk geometry is not AdS, and the dual theory is not
conformal; instead the bulk describes the dual of a non-trivial flow  
between two different fixed points characterized by a physical,
dynamically-generated scale $\Lambda_{\ast}$.  
Second, we do not attribute any physical meaning to the two
boundaries, and hence we are not interested in   
fine-tuning considerations involving the choices of parameters in the
boundary actions. 
Finally, the system of scalars we write in the bulk 
is not dual to a quasi-marginal deformation, but rather to 
relevant operators that acquire large anomalous dimensions, 
and a resulting RG flow containing non-trivial 
operator-mixing effects.
In practice, the above means that we are only interested in 
the physical effects due to those elements of the five-dmensional
supergravity sigma-model that have a fully dynamical physical
origin. The presence of the two boundaries is just a technical device,
and we 
would  remove them completely from the analysis if a complete flow
yielding confinement in the IR were known.

\section{Discussion}
\label{sec:discuss}

In this paper we have studied various aspects of the Leigh-Strassler flow and its modifications, with a view towards constructing a holographic model of quasi-conformal dynamics (walking), within a consistent supergravity/string
theory framework. We have seen that a combination of the Pilch-Warner
supergravity solution and a hard-wall IR cut-off exhibits a light
dilaton-like state in the spectrum. We have also seen that the flow
towards the strongly interacting Leigh-Strassler CFT generates large
(negative) anomalous 
dimensions for chiral operators and for mesonic operators built from excitations of a flavor D7-brane in the geometry. Both of these are positive results from the viewpoint of using this flow to model
WTC dynamics. In particular, this encourages us to look for a complete
description of the flow including the departure from the IR
fixed-point region to a confining vacuum within the
supergravity/string framework. 

Perhaps the most important question that we would like to answer in
this framework  is under what general conditions will a light
pseudo-dilaton persist in the spectrum. The appearance of a 
pseudo-dilaton is associated to spontaneous breaking of
approximate conformal invariance. This means that what drives the
theory away from the IR fixed point is primarily a condensate for some
operator, and this must be a guiding criterion in the search for a
complete description of the flow. 

\subsection{Towards IR completions}

%{\bf Towards IR completions:} 
In the hard-wall picture, conformal invariance is spontaneously
broken by the IR cut-off and the details of which field theory
operators condense cannot be addressed. To actually understand this,
as we have explained in Section \ref{intro}, we need to study the
deformation of the Pilch-Warner flow by operators that make the field 
theory confine in the deep IR. The operators should be those that
remain relevant or marginally relevant at the IR fixed
point. In principle, these include supersymmetric mass deformations, 
certain cubic and quartic superpotential deformations of the ${\cal
  N}=4$ theory, or even non-supersymmetric deformations. For the
${\cal N}=1^*$ mass deformation, approximate IIB string 
backgrounds were constructed by Polchinski and Strassler \cite{PS}
(for equal masses), 
and the dynamics of IR confinement is described by 5-brane configurations
that appear in the interior of the geometry. To explore the theory with a
walking regime $(m_1=m_2 =m \ll m_3)$, a further deformation of this
background will be necessary\footnote{The 5-branes in the IR wrap flux-supported two-cycles. In
  the walking regime these 2-cycles will resemble flattened discs (on
  which D3-charge is smeared) which may possibly lead to
  simplification. It is not clear that the approximations employed in
  \cite{PS} will hold in this regime.}
 (see Appendix \ref{app:psdef}).

For the specific case of the ${\cal N}=1^*$ deformation, much can be
said based on field theory considerations alone.
The magnitudes of all chiral condensates can be exactly computed
\cite{dorey,doreykum, adk, mm} as functions of the masses $m_i$ and the
gauge coupling. For example,
in the confining phase  it can be shown that 
\be
\langle\Tr
\Phi_3^2\rangle \sim {N}\, m^2 \,E_2\left(\frac{i}{\lambda}\right) \sac 
\langle\Tr\Phi_{1,2}^2\rangle \sim N \, m_3 m
\,E_2\left(\frac{i}{\lambda}\right) \,,
\ee
where $\lambda\equiv{g^2_\mt{YM} N}/{4\pi}$ is the 't Hooft
coupling and $E_2$ is the second
Eisenstein series, a quasi-modular form. 
Similarly, the gluino condensate depends on
the combination $N m^2
m_3\,E_2^\prime\left(\frac{i}{\lambda}\right)$. All the quantities
interpolate 
between weak coupling and 
the limit of large 't Hooft coupling, corresponding to
the dual supergravity limit. At strong 't Hooft coupling we expect
that the dynamical scale of confinement (set by the gluino condensate,
for example) will be $\Lambda_0\sim (m^2 m_3)^{\frac
  {1}{3}}$, whilst the largest condensates (after factoring out their
$N$-dependence)  are set by the scale $\sqrt{m m_3}$,  which is
parametrically larger in the `walking limit'  $m_3\gg m$.  
This fact suggests that the ${\cal N}=1^*$ flow near the LS fixed point will
be forced to deviate by large VEVS for chiral operators, which is
precisely the type of situation in which we expect an approximate
dilaton-like state.  

A useful first step towards the construction of a string dual picture for
the deformation of the Pilch-Warner flow would be to study the dual gravity description of a flow that terminates at a point on the Coulomb
branch of the IR CFT \cite{Gowdigere:2005wq}.
Such geometries will be sourced by continuous
distributions of D3-branes in the IR, encoding the VEVs on the Coulomb
branch. Importantly, the flow towards the fixed point will deviate due
to VEVs alone and be cut off by the D3-brane distribution. 
It would be interesting to understand how to extract the Goldstone
mode of spontaneously broken dilatation invariance in the IR theory, from
such a SUGRA background.  The corresponding 
analysis of the spectrum from the gravity dual viewpoint, for 
Coulomb branch configurations of ${\cal N}=4$ SYM was perfomed
in \cite{Brandhuber:2000fr, Bianchi:2000sm}.

\subsection{Deformations of D7-brane embeddings}

Another direction that would be interesting to explore, to look for a light scalar, lies within the probe-brane sector we have studied in this paper.  In this context we would keep the gravitational
background (Pilch-Warner) unchanged while
having conformal symmetry broken by the embedding of the $N_f$ probe
D7-branes. In particular, the conformal symmetry
 is broken in the sector of fundamental matter while the glue theory remains conformal
(which is only possible in the 't Hooft limit $N_f / N \to 0$).

It would then be natural to look for situations in which the fermion bilinear $q^i {\tilde q}_i$ condenses, namely, 
the `slipping mode'  of the D7-brane gets a non-trivial
vacuum expectation value. This is only possible if supersymmetry is broken. Starting
from the ${\cal N}=4$ UV fixed point, one could
imagine tilting the D7-brane embedding 
in such a way that the supersymmetries broken by the fundamental matter fields
are incompatible with those broken by the mass deformation $m_3$. 
Put another way, neither the adjoint mass deformation nor the D7-brane
alone would break all the supersymmetries on their own, but the combination of both could. This can
be achieved without adding an explicit mass for the fundamental
flavors and is allowed
since the global symmetry of ${\cal N}=4$ theory is $SU(4)$, which is
larger than the manifest $SU(3)$ that
rotates the three chiral adjoints once their holomorphic structure is fixed. The classical Lagrangian
will still be conformally invariant but it is plausible that the
slipping mode condenses and spontaneously breaks conformal symmetry of
the IR fixed point (of the probe sector).

It would also be interesting to find (possibly non-supersymmetric)
D7-brane embeddings which may spontaneously break  
the conformal symmetry of either the UV or the IR fixed point, in analogy to what was found
in \cite{Kuperstein:2008cq} in a different framework. It would then be
very interesting to see how these  D7-branes are affected by the flow
and to understand how the masses of the Goldstone bosons may be
lifted.  We would also like to mention that seven-branes in the
walking geometry of \cite{NPP} have been analysed 
in \cite{Anguelova:2010qh}.

%%%%%%%%%%%%%%%%%%%%%%%%%%%%%%%%%%%%%%%%%%%%%%%%%%%%%%%%%%%%%%%%%%%%%%
%% Acknowledgments %%%%%%%%%%%%%%%%%%%%%%%%%%%%%%%%%%%%%%%%%%%%%%%%%%%
%%%%%%%%%%%%%%%%%%%%%%%%%%%%%%%%%%%%%%%%%%%%%%%%%%%%%%%%%%%%%%%%%%%%%%
%\vspace{1.0cm}
\begin{acknowledgments}
It is a pleasure to thank Roberto Emparan, Jaume Garriga, Tim Hollowood and Carlos N\'u\~nez for helpful discussions.
The work of DM and AP is supported by grants FPA2007-66665C02-02 and DURSI 2009 SGR 168, and by the CPAN CSD2007-00042 project of the Consolider-Ingenio
2010 program. DM is also supported by grant FPA2007-66665C02-01. The work of MP is supported in part  by WIMCS and by the STFC grant ST/G000506/1. SPK acknowledges support from STFC grant ST/G000506/1.

\end{acknowledgments}

%\newpage
%\startappendix
\appendix

\section{Relating Pilch-Warner to ${\cal N}=1^*$}
\label{app:psdef}

The gravity dual of ${\cal N}=1^*$ was discussed by 
Polchinski and Strassler in \cite{PS}.
It is obtained by deforming ${\cal N}=4$ SYM by masses of the
adjoint multiplets $\Phi_i$ ($i$=1,2,3). 
At first order,
the deformation amounts to turning on the 3-form. Using notation
of \cite{PS}:
\be
T_3 = m_1 dz^1 \wedge d\bar z^2 \wedge d\bar z^3 + 
m_2 d\bar z^1 \wedge d z^2 \wedge d\bar z^3 + 
m_3 d\bar z^1 \wedge d\bar z^2 \wedge dz^3 + 
m_4 dz^1 \wedge d z^2 \wedge dz^3 
\label{T3PS}
\ee
where we have included $m_4$, namely a (supersymmetry breaking) gluino mass.
$T_3$ is proportional to the complex 3-form. 
The deformation leading the IR Leigh-Strassler fixed point
 corresponds to taking a mass for just one of the adjoint
multiplets, namely $m_1=m_2=m_4=0$ in (\ref{T3PS}). We can then write:
\be
S_2|_{m_1=m_2=m_4=0} = m_3 \left( \bar z^1 \wedge d\bar z^2 \wedge dz^3 + \bar z^2 \wedge dz^3 \wedge d\bar z^1 + 
z^3 d\bar z^1 \wedge d\bar z^2 \right)
\label{S2PW}
\ee
such that $dS_2 = 3 T_3$.
We now check that, near the UV fixed point, the two-form of the Pilch-Warner
solution written in (\ref{C2B2}) is proportional to this quantity.
Using $\alpha \approx 0$, $\tanh \chi  \approx m_3\, e^{-r}$, $X_1 \approx 1$,
we find the behaviour of (\ref{C2B2}) in the asymptotic UV:
\be
C_2 + iB_2|_{PW}\approx\frac{m_3 \, e^{-r}}{2}
e^{-i\,\varphi} (\cos \theta d\theta +\frac{i}{2}\,\cos^2\theta \sin\theta\, \sigma_3 -
i\, \cos^2\theta \sin\theta d\varphi)\wedge (\sigma_1 - i\,\sigma_2)\,\,,
\label{A2express}
\ee
Making contact between (\ref{S2PW}) and (\ref{A2express}) 
requires writing the complex coordinates in
terms of the angles. 
Near the UV, we can write:
\be
z_1 = e^r \cos \theta \,\cos\frac{\alpha_1}{2} e^{\frac{i}{2}(\alpha_3 + \alpha_2)} \,\,,\ \
z_2 = -e^r\cos \theta \,\sin\frac{\alpha_1}{2} e^{\frac{i}{2}(\alpha_3 - \alpha_2)} \,\,,\ \
z_3 = e^r e^{-i\,\varphi} \sin\theta\,\,.
\label{zsexplicit}
\ee
It is now straightforward to check that 
$S_2|_{m_1=m_2=m_4=0} \approx e^{4r} (C_2 + iB_2)|_{PW}$.

With the above identifications,
one can try to deform the PW flow with by incorporating small
masses $m_1,m_2$ or $m_4$. It is worth mentioning that in the case of \cite{PS}, the
deformation of ${\cal N}=4$ SYM at first order in the $m_i$ only includes the 3-form --- for instance,
the dilaton is excited only at quadratic order. Conversely, 
if one deforms the Pilch-Warner flow with, say, a small $m_1$, the
dilaton is already excited at linear order in the deformation. 
This happens because
in $F_3^2 - H_3^2$ there would be terms proportional to $m_1\,m_3$ which cannot be neglected at first order
since $m_3$ is not small in this case.

\section{Singular flows}
\label{app:sing}

In this appendix we discuss the behaviour of the classical solutions 
to the BPS system in Eqs.~(\ref{eqforA}) and~(\ref{chialphaeqs}),
in proximity of the end-points of the flows, labelled $S_{\pm}$ and $S_I$ in Fig.~\ref{Fig:RG}.
In particular, we briefly discuss the physics inferred from probing the resulting geometries with strings
dual to Wilson-loop operators.
The asymptotic solution near $S_U$ is not useful for our purposes, and hence it is omitted here.

We begin by noticing that the BPS equations admit a further truncation, by fixing $\chi=0$.
In this case, one is perturbing the UV fixed point by introducing mass terms and VEVs for the scalar, 
but not for the fermion.
The resulting equation is
\beqs
\partial_{r} \alpha&=&
-\frac{2}{3}e^{\alpha}\sinh 3 \alpha\,,
\label{singularaeq}
\eeqs
which admits two possible asymptotic solutions in the IR, depending on the sign of $\alpha$.
Expanding the solution with positive $\alpha$ (the $S_+$ point in Fig.~\ref{Fig:RG}), 
and fixing an integration constant so that the singularity appears at $r_0=0$, 
at leading order in $r$ one has
\beqs
e^{\alpha}&=&\left(\frac{4r}{3}\right)^{-\frac{1}{4}}\,+\,\frac{2 \sqrt{2}}{15\, 3^{1/4}}r^{5/4}\,+\cdots\,,\\
e^{A}&=&r^{\frac{1}{4}}\,+\,\frac{16 r^{7/4}}{15\, \sqrt{3}}\,+\cdots\,.
\eeqs
By inspection of the classical equations derived from the Nambu-Goto action with this background, 
we find that there is no way to take $L_{QQ}\rightarrow +\infty$, 
in the sense that the classical system  does not admit any solution
with arbitrary large separation 
between the quak-antiquark pair. This cannot hence be interpreted as the dual of a confining theory

The solution to (\ref{singularaeq})
 with negative $\alpha$ (the $S_-$ point in Fig.~\ref{Fig:RG}) yields similar results.
Notice however (see again Fig.~\ref{Fig:RG}) that the case of negative $\alpha$ is 
 unstable, in the sense that even a small value of $\chi$ drives the flow towards the
fixed point with $\chi\rightarrow +\infty$ (the point $S_I$). 
Also, notice that there are no walking flows ending near $S_-$, 
hence we are not really interested in this type of solution.

Reinstating $\chi\neq 0$ in the equations, and choosing one
integration constant so that the singularity appears at ${r_0}=0$, 
the  asymptotic solution  around the point $S_I$  can be expanded as
\beqs
e^{2\alpha}&=&a_0r^{1/7}+{\cal O}(r^{4/7})+{\cal O}(r)\,,\\
e^{2\chi}&=&\frac{6 a_0}{7}r^{-6/7}+{\cal O}(r^{-3/7})+{\cal O}(1)\,,
\eeqs
Fixing the second integration constant  $a_0=7/6$ for convenience, we write:
\beqs
e^{2A}&=&c_0r^{2/7}+{\cal O}(r^{5/7})\,,
\eeqs
We may also fix the third integration constant to $c_0=1$, which is irrelevant for the upcoming discussion.
A  careful analysis shows that $L_{QQ}$ is actually finite even when
the string tip approaches the singularity ($\hat r_0\rightarrow 0$), which again means that one cannot take
$L_{QQ}\rightarrow +\infty$, and hence this is not interpreted as the dual description of a confining theory. 
The situation is very similar to that discussed in~\cite{Apreda:2003sy}. 

In summary, we conclude that none of the possible flows ends with a healthy behaviour that can be interpreted
as a signal of confinement of the dual theory
(even though $S_I$ might yield the best toy model in that respect), 
and hence a radical modification of the IR geometry is needed in this respect

\section{Scalar and vector harmonics in the Berger sphere}
\label{app:harmonics}

In section \ref{sec:D7}, we discussed the physics of a D7-brane living
in the IR fixed 
point of the Pilch-Warner solution. The worldvolume of the D7 is
$AdS_5 \times \tilde S^3$, 
where $\tilde S^3$ is the so-called Berger sphere or squashed sphere.
For the analysis of section \ref{sec:D7}, it is
important to know the harmonics on $\tilde S^3$, which we discuss
in this appendix.
In particular, we are interested in finding the eigenvalues of the
Hodge-de Rham operator for scalars and vectors. The Hodge-de Rham
operator acting on 
a $p$-form is defined, using notation of \cite{Duff:1986hr}, 
 as $\Delta = (-1)^p (d*d* + *d*d)$, where $d$ is the usual exterior derivative
and the star represents Hodge duality.

Let us start by writing the metric of $\tilde S^3$ in terms of two constants.
\be
d\tilde s_3^2 = a^2 (\sigma_1^2 + \sigma_2^2) + b^2 \sigma_3^2 = 
a^2(d\alpha_1^2 + \sin^2 \alpha_1 d\alpha_2^2) + b^2 (d\alpha_3 + \cos
\alpha_1 d \alpha_2)^2\,\,, 
\label{metrics3}
\ee
such that the unit round $S^3$ is recovered with $a^2 = b^2 = \frac14$ and the metric
of (\ref{squsS3}) has $a^2 = 1$, $b^2 = \frac43$.
It is useful to introduce a coordinate $z$ defined as:
\be 
z= \cos \alpha_1
\ee
The ranges of the different coordinates are $z\in [-1,1]$, $\alpha_2 \in [0,2\pi)$,
$\alpha_3 \in [0,4\pi)$. In the following, we will make use of the
so-called Wigner functions or $SU(2)$ functions, which
we denote as $P^l_{mn}(z)$, following \cite{vilenkin}, where their
 properties are discussed at length.
 \bear
P^l_{mn}(z)=\frac{(-1)^{l-n}i^{n-m}}{2^l}\sqrt{\frac{(l+m)!}{(l-n)!(l+n)!(l-m)!}}\times\rc
(1+z)^{\frac{-m-n}{2}}(1-z)^{\frac{n-m}{2}}
\frac{d^{l-m}}{dz^{l-m}}\left[
(1-z)^{l-n} (1+z)^{l+n} 
\right]\,\,,
\label{plmn}
\eear
where the allowed values of $l,m,n$ are
\be
l=0,\frac12,1,\frac32, \dots , \infty\,,\qquad\
m= -l , -l+1,\dots,l-1,l\,,
\qquad\
n= -l , -l+1,\dots,l-1,l\,.
\ee
The $P^l_{mn}(z)$ are generalizations of the Legendre polynomials and associated
Legendre polynomials in the sense that $P_l(z) = P^l_{00}$ and
$P_l^m(z)$ is a constant number times $P^l_{m0}(z)$ (see \cite{vilenkin} for details).
They solve the following second order differential equation:
\be
\left[ (1-z^2)\, \partial_{zz} -2z\,\partial_z
- \frac{m^2 -2mnz +n^2}{1-z^2}\right] P^l_{mn}(z) = -l(l+1) P^l_{mn}(z)
\label{eqforPlmn}
\ee

\subsection*{Scalar harmonics}

The scalar harmonics of the $\tilde S^3$ together with some discussion of
the geometry of the squashed sphere can be found in appendix A of \cite{Zoubos:2004qm}.
The Hodge-de Rham operator acting on a scalar is just minus the laplacian. 
The eigenfunctions are given by:
\be
Y^l_{mn} = e^{-i\,m\,\alpha_2-i\,n\,\alpha_3} P^l_{mn}(z)\,\,,
\ee
from where, using (\ref{eqforPlmn}) we  find:
\be
\Delta Y_{mn}^l =-
\nabla^2  Y^l_{mn} =- \frac{1}{\sqrt {\det [g]}} \partial_i \left(\sqrt {\det [g]}
g^{ij} \partial_j Y^l_{mn}
\right) = \lambda\, Y^l_{mn}
\label{laplacian}
\ee
such that the eigenvalues are:
\be
\lambda = \left(\frac{l(l+1)}{a^2}+ \left( \frac{1}{b^2} - \frac{1}{a^2}\right) n^2 \right)
\label{lambdascalar}
\ee

\subsection*{Vector harmonics}

We have not been able to find in the literature a discussion of vector harmonics for the
Berger sphere, and we present them here. The 
Hodge-de Rham operator acting on a vector can be written as \cite{Duff:1986hr}
$\Delta f_i = - g^{jk} \nabla_j \nabla_k f_i + R_i^j f_j$ where
$\nabla$ is the usual covariant
derivative and $R_m^n$ is the Ricci tensor. 
It can be rewritten as
\be
\Delta f_i = - g^{jk} \nabla_j \nabla_k f_i + R_i^j f_j = 
{\cal O}_1 ( {\cal O}_1 f_i) + {\cal O}_2 f_i\,\,,
\label{HdRvectors}
\ee
where we have introduced the differential operators:
\be
{\cal O}_1 f_i= \frac{ g_{ip}}{\sqrt{\det [g]}}\epsilon^{pjk} \partial_j  f_k 
\,\,,\qquad
{\cal O}_2 f_i =-
 \partial_i \left( \frac{1}{\sqrt{ \det [g]}}
 \partial_k(
\sqrt{ \det [g]} \, g^{kj}  f_j)\right)
\label{O1O2}
\ee
We now look for eigenvectors of (\ref{HdRvectors}), namely
\be
\Delta f_i = \lambda_v f_i
\ee
The simplest solution to this equation is built by taking the gradient of the
scalar harmonics, namely
\be
\left({\bf Y}^l_{mn}\right)_i =  \partial_i Y^l_{mn}\,\,,\qquad
l=\frac12,1,\dots
\label{trivialvectorharm}
\ee
The eigenvalues are again given by (\ref{lambdascalar}). Notice that $l=0$ is not allowed,
since $\partial_i Y^0_{00}=0$.

The rest of vector harmonics are killed by
${\cal O}_2$ whereas they are eigenvectors of ${\cal O}_1$. 
There are two families of such solutions which we denote by $\pm$.
Let us write
\be
{\bf Y}^\pm = e^{-i\,m\,\alpha_2-i\,n\,\alpha_3}
\left(Y_1^{\pm} dz + Y_2^{\pm} d\alpha_2 + Y_3 d\alpha_3\right)
\label{vectorharm1}
\ee
where, explicitly:
\bear
Y_3(z) &=& P^l_{mn}(z) \,\,,\rc
Y_2^{\pm}(z) &=& \frac{(-m\,n  + b^2 v_\pm^2 z)Y_3(z)-
b\,v_\pm\,(1-z^2)Y_3'(z)}{b^2 v_\pm^2 - n^2} \,\,,\rc
Y_1^\pm(z)&=&  \frac{i\,(n\,Y_2^\pm(z) - m\, Y_3(z))}{b\,v_{\pm}\,(1-z^2)}
\label{explicitYs}
\eear
and the eigenvalues of the  ${\cal O}_1$ operator
\be
{\cal O}_1 {\bf Y}^\pm=v_{\pm} {\bf Y}^\pm
\label{eigenO1}
\ee
 are given by:
\be
v_\pm = \frac{b^2 \pm \sqrt{b^4 + 4 a^4 n^2 + 4 a^2 b^2 (l^2 + l -n^2)} }{2 a^2 b}
\label{vpm}
\ee
The  $\lambda_v$ are found by squaring these expressions.
One can check that, for the $-$ mode, when $|n|=l$, the denominator in the expression for
$Y_2^-$ vanishes, and there is no regular solution. Thus, the minus modes exist only for
$|n| < l$. On the other hand, there is an extra solution not included in 
(\ref{explicitYs}), found by setting $Y_3=0$. It can be thought of as the $n=\pm(l+1)$ generalization of the
$+$ mode described above. It reads\footnote{In order to check that (\ref{extraplus}) solves
(\ref{eigenO1}), one can use the following identities \cite{vilenkin}
\begin{displaymath}
P^l_{mn}(z)=P^l_{nm}(z)\,\,,\qquad
P^l_{mn}(-z)=i^{2l-2m-2n}P^l_{m,-n}(z)\,\,,\qquad
\end{displaymath}
and the definition (\ref{plmn})
to prove that $\frac{d}{dz}P^l_{m,\pm l}(z)=\frac{\pm m-l\,z}{1-z^2}P^l_{m,\pm l}(z)$.
}:
\be
{\bf Y}^+|_{n=\pm(l+1)}= e^{-i\,m\,\alpha_2 \mp i\,(l+1)\,\alpha_3} P^{l+1}_{m,\pm(l+1)}
\left(\frac{\pm i}{1-z^2} dz +  d\alpha_2 \right)\,\,,
\label{extraplus}
\ee
which satisfies ${\cal O}_1 {\bf Y}^+|_{n=\pm(l+1)} = (l+1)/b\ {\bf Y}^+|_{n=\pm(l+1)}$.
Summarizing, the spectrum of the ${\cal O}_1$ operator reads
\bear
v_+ &=& \frac{b^2 + \sqrt{b^4 + 4 a^4 n^2 + 4 a^2 b^2 (l^2 + l -n^2)} }{2 a^2 b}\,,\qquad
\left(
|n|\leq l+1 ,\quad l=0,\frac12, \dots\right)
\label{vp} \\
v_- &=& \frac{b^2 - \sqrt{b^4 + 4 a^4 n^2 + 4 a^2 b^2 (l^2 + l -n^2)} }{2 a^2 b}\,,\qquad
\left(
|n|\leq l-1 ,\quad l= 1,\frac32 \dots\right)
\label{vm}
\eear
with  $m= -l,\dots, l$ in all cases. 
It is interesting that, for $|n|=l$ and $|n|=l+1$ the square root in $v_+$ can be explicitly
performed and we find
\be
v_+|_{|n|=l+1}= \frac{l+1}{b}\,\,,\qquad \qquad
v_+|_{|n|=l}= \frac{b}{a^2}+\frac{l}{b}\,\,.
\label{vpspecial}
\ee
As a final note, let us remark that,
using the properties of the $P^l_{mn}(z)$, one can check that the harmonics presented are
regular on the whole sphere. The easiest way to check it is to rewrite them in terms of a
dreibein of (\ref{metrics3}) and make use of the fact that the leading
behaviour of the 
Wigner function
 around $z=1$ is $P^l_{mn}(z) \sim (1-z)^{|m-n|/2}$ --- apart from a multiplicative constant ---
and around $z=-1$ is $P^l_{mn}(z) \sim(1+z)^{|m+n|/2}$.

\section{A simplified example: Small deviations from AdS}
\label{simple}
In order to understand the numerical results presented in this
Section \ref{Sec:3}, it is useful to perform a simplified exercise. We consider the case in which we are given a fully back-reacted 5D
model, wherein 
the sigma-model has trivial kinetic terms, and the superpotential can
be approximated  
as a quadratic function of the scalars. We also assume that the two
boundaries $r_1$ and $r_2$ 
are chosen in such a way that effectively the whole space between them
has a geometry that is close to AdS. In practice, this 
means that we require the two scalar sigma-model VEVs to have small
values at both boundaries (and  
as a consequence in the whole bulk).
We apply to this situation, a slight generalization of the 
technique used in~\cite{EP} to the case of the two scalars here. 
%generalizing to
%the case of two scalars what  
%done there for one scalar with quadratic superpotential, 
We pay particular attention to the calculation of the 
masses of the two lightest states in the two towers corresponding to
the sigma-model scalars. 

We begin by writing the superpotential of the canonically normalized
$X^a$ ($a=+,-$) scalars as  
\beqs
W&=&-\frac{3}{2}-\frac{1}{2}X^T \Delta X\,,
\eeqs
where $\Delta={\rm diag}\,\{\Delta_+,\Delta_-\}$ is assumed to be diagonal without loss of generailty. 
The BPS equations are solved by
\beqs
X_+&=&X_+(r_2)e^{-\Delta_+(r-r_2)}\,,\\
X_-&=&X_-(r_2)e^{-\Delta_-(r-r_2)}\,,\\
A&=&r-\frac{1}{6}\left(X_+(r_2)^2e^{-2\Delta_+(r-r_2)}+X_-(r_2)^2e^{-2\Delta_-(r-r_2)}\right)\,,
\eeqs
where an integration constant has been chosen so that for vanishing VEVs one recovers $A = r$.

Because of our assumptions we have that $X_{\pm}(r_{1,2})\ll 1$, which in turn 
implies that the equations for the fluctuations can be solved 
perturbatively by expanding in powers of the boundary VEVs.
As explained in~\cite{EP}, to look for the lightest solutions
one needs to solve the bulk equation
(\ref{Eq:diffeqWN}) after setting $q^2=0$ and all the boundary VEVs to zero (to leading order in the small VEVs), 
subject to the boundary conditions Eq.~(\ref{Eq:BCWinfty}) in which
$W$, $A$ and $N$ are evaluated by setting the VEVs to zero
(leading-order). However in $W^c$, one keeps the leading-order
dependence on the VEVs, which will then show up in  
$q^2$ as well, ensuring that all terms in all equations are written at
the same (leading) order. 

The bulk equations decouple from each other in this limit, and read simply
\beqs
   \Bigg[  \left( \partial_r - \Delta_+ +4\right) \left( \partial_r + \Delta_+ \right)  \Bigg] \mathfrak{a}_+&=&0\,,\\
      \Bigg[  \left( \partial_r - \Delta_- +4\right) \left( \partial_r + \Delta_- \right)  \Bigg] \mathfrak{a}_-&=&0\,,
\eeqs
which are solved by
\beqs
\mathfrak{a}_+&=&c_1^{+}e^{-\Delta_+r}\,+\,c_2^{+}e^{-(4-\Delta_+)r}\,,\\
\mathfrak{a}_-&=&c_1^{-}e^{-\Delta_-r}\,+\,c_2^{-}e^{-(4-\Delta_-)r}\,.
\eeqs

Note that, although the superpotential seems to
suggest that the scalars decouple from each other, this is not the case:  the boundary terms form a coupled system. 
However, at this point the system reduces to a set of four  algebraic equations in the variables $q^2$ and $c_i^{\pm}$,
with one arbitrary overall normalization constant:
\begin{displaymath}
%\left.
-\frac{2}{3}\frac{e^{2r}}{q^2}
\left(\begin{array}{cc}
\Delta_+^2X_+(r)^2 &\Delta_+X_+(r) \Delta_-X_-(r) \cr
\Delta_+X_+(r) \Delta_-X_-(r) & \Delta_-^2X_-(r)^2 \end{array}\right)
\left(\partial_r+
\left(\begin{array}{cc}
\Delta_+ &\cr
&\Delta_- \end{array}\right)\right)
\mathfrak{a}\Bigg|_{r_i}=
\frac{}{}\mathfrak{a}\Bigg|_{r_i}\,.
\end{displaymath}
This  yields
\bear
q^2&=&\frac{4 (\Delta_+-2) \Delta_+^2 e^{(\Delta_-+5)
   r_1+(\Delta_-+2 \Delta_++5) r_2} X_+(r_2)^2
    \frac{\sinh
   (r_1-r_2)}{\sinh((\Delta_+-2) (r_1-r_2))}}
   {3 (\cosh ((\Delta_-+\Delta_++4)
   (r_1+r_2))+\sinh ((\Delta_-+\Delta_++4)
   (r_1+r_2)))}+\,(+\leftrightarrow -)\,.\nonumber
\eear

Let us apply all of this to our example.
We begin by assuming that $r_I\ll r_1 \ll r_2 \ll r_{\ast}$, so that the 
slice of geometry is well approximated by the IR fixed-point.
From \eqn{deltapm} we have $\Delta_{\pm}=1\pm\sqrt{7}$.
After some algebra we find
\beqs
q^2&=&\frac{8 (-1 + \sqrt{7}) e^{
 2 (\sqrt{7}r_1 + r_2)} (-e^{2 r_1} + e^{2 r_2}) X_-(r_2)^2}{-e^{
  2 (1 + \sqrt{7}) r_1} + e^{2 (1 + \sqrt{7}) r_2}}+\\ \nonumber
 &&+ \frac{4 (1 + \sqrt{7}) e^{-2 (r_1 + \sqrt{7} r_1 + 2 r_2)} (-e^{2 r_1} + e^{
   2 r_2}) (-e^{2 (3 + \sqrt{7}) r_2} + e^{
   2 (r_1 + \sqrt{7} r_1 + 2 r_2)}) X_+(r_2)^2}{
\cosh 2 (r_1 - r_2)]- \cosh 2 \sqrt{7}(r_1 - r_2)}
  \,,\\
  &\propto&e^{2(1-\sqrt{7})r_2}\,,
\label{eq:massformula}
\eeqs
where in the last line we have used the fact that $r_2\gg r_1$ and the 
replacement $X_+(r_2)=X_+(r_1)e^{-(1 + \sqrt{7}) (r_2 - r_1)}$.
Which is to say that for a given (small) value of the VEV of the operator of dimension
$1+\sqrt{7}$ (measured in the IR) and given (small) value for the coupling of the irrelevant operator of
dimension $3+\sqrt{7}$ (measured in the UV), there is a parametrically light scalar,
with exponentially suppressed mass $\propto e^{(1 - \sqrt{7}) r_2}$
as a function of the UV cut-off.

The same exercise can be carried out in the other extreme case in which
$r_{\ast}\ll r_1\ll r_2$, and hence the whole geometry is close to the UV fixed point,
 in which case $\Delta_{\pm}=3/2\pm1/2$. In this case,
\beqs
q^2&=&\frac{4 e^{2 r_2-2 r_1}}
{3  (r_2-r_1)}
\Big(-e^{4 r_1-2r_2} r_1
   X_-(r_1)^2+e^{4 r_1-2r_2} r_2 X_-(r_1)^2-2 e^{6
   r_1-4r_2} X_+(r_1)^2+
   \rc
   &+&
   2 e^{-2 r_2+4r_1} X_+(r_1)^2\Big)
   \simeq
   \frac{4}{3}e^{2r_1}\left( X_-(r_1)^2+\frac{2}{r_2} X_+(r_1)^2\right)\,,
\eeqs
where again in the last expression we use the fact that $r_2\gg r_1$.
Notice that in this case one of the two deformations is the insertion
of a relevant operator, 
and hence there is no suppression due to $r_2$ (the UV cut-off). 
The other deformation is the VEV of a dimension-two  
operator, and hence there is a quasi-marginal double-trace deformation
producing an $r_2$-suppressed mass 
proportional to $X_+(r_1)$. However, the quasi-marginal nature of the
latter means that the suppression is not exponential 
(recall that the coordinate $r$ 
is related to the logarithm of  the physical scale in canonical units in the dual field theory). 
Notice also that in both examples we applied the rule (dictated by common sense) that we express  the mass 
as a function of the VEV of the fields $X_{\pm}$, by evaluating at $r_1$ those that would drive the flow
away from the fixed point in the IR, while evaluating at $r_2$ those that would drive away the flow towards the UV.

We can perform a similar exercise in order to determine the masses of the heavy states.
Now we are considering states whose mass is not suppressed by the VEVs of the sigma-model scalars.
Hence we need to keep the $q^2$-dependence in the bulk equations, while the boundary conditions
become Dirichlet:
\beqs
   \Bigg[  \left( \partial_r - \Delta_+ +4\right) \left( \partial_r + \Delta_+ \right)  + q^2 e^{-2r}\Bigg] \mathfrak{a}_+&=&0\,,\\
      \Bigg[  \left( \partial_r - \Delta_- +4\right) \left( \partial_r + \Delta_- \right)   + q^2 e^{-2r} \Bigg] \mathfrak{a}_-&=&0\,,\\
      \mathfrak{a}\Bigg|_{r_i}&=&0\,.
\eeqs
The solutions can be written in terms of Bessel functions $J$ and $Y$:
\beqs
 \mathfrak{a}_+&=&e^{-2r}\left[c_1^{+}J_{2-\Delta_+}(e^{-r}q)+c_2^{+}Y_{2-\Delta_+}(e^{-r}q)\right]\,,\\
  \mathfrak{a}_-&=&e^{-2r}\left[c_1^{-}J_{2-\Delta_-}(e^{-r}q)+c_2^{-}Y_{2-\Delta_-}(e^{-r}q)\right]\,,
  \eeqs
and therefore one can easily obtain the entire spectrum.

For the case where the theory is always close to the UV fixed point, one gets for $n=1,2,\cdots$:
\beqs
M_+&\simeq&\pi e^{r_1}\left(\frac{3}{4}+n-1\right)\,,\\
M_-&\simeq&\pi e^{r_1}\left(\frac{5}{4}+n-1\right)\,.
\eeqs
For the case where the theory is always close to the IR fixed point, one gets for $n=1,2,\cdots$:
\beqs
M_+&\simeq&\pi e^{r_1}\left(-\frac{\sqrt{3}}{4}+n+1\right)\,,\\
M_-&\simeq&\pi e^{r_1}\left(\frac{\sqrt{3}}{4}+n+1\right)\,,
\eeqs
which fits the numerics  very well.
Notice in particular that, due to the numerical coincidence $\sqrt{3}/2\sim 0.9$, the two towers
appear to be almost degenerate. The exceptions to this are  the
first light states, where the approximation 
is somewhat bad, and most importantly the 
 lightest state of the $M_+$ series which is not paired to any state in the $M_-$ tower.
There is no deep symmetry reason for this.

\end{document}